\documentclass[12pt]{article}
\pdfoutput=1

\usepackage{amsmath,amssymb,amsfonts,epsfig,cite,setspace,bigstrut,framed,
float,adjustbox,hyperref}
\usepackage[all]{xy}
\usepackage{color}
\usepackage{pifont}
\usepackage{ulem}
\usepackage{marginnote}
\hypersetup{
    colorlinks=true,
    linkcolor=black,
    citecolor=blue,
    urlcolor=blue,
    linktocpage}

\makeatletter \@addtoreset{equation}{section} \makeatother
\renewcommand{\theequation}{\thesection.\arabic{equation}}

\begin{document}

\vskip 0.25in

\newcommand{\todo}[1]{{\bf\color{blue} !! #1 !!}\marginpar{\color{blue}$\Longleftarrow$}}
\newcommand{\nn}{\nonumber}
\newcommand{\comment}[1]{}
\newcommand\T{\rule{0pt}{2.6ex}}
\newcommand\B{\rule[-1.2ex]{0pt}{0pt}}

\newcommand{\tocite}{{\textcolor{blue}{[!CITE!]}}\marginpar{$\Longleftarrow\Longleftarrow!!!$}}
\newcommand{\tocheck}{\marginpar{$\Longleftarrow\Longleftarrow!!!$}}
\newcommand{\towrite}{{[WRITING...]}\marginpar{$\Longleftarrow\Longleftarrow!!!$}}

\newcommand{\cG}{{\cal G}}
\newcommand{\cV}{{\cal V}}
\newcommand{\cP}{{\cal P}}
\newcommand{\cL}{{\cal L}}
\newcommand{\cO}{{\cal O}}
\newcommand{\cI}{{\cal I}}
\newcommand{\cM}{{\cal M}}
\newcommand{\cW}{{\cal W}}
\newcommand{\cN}{{\cal N}}
\newcommand{\cR}{{\cal R}}
\newcommand{\cH}{{\cal H}}
\newcommand{\cK}{{\cal K}}
\newcommand{\cT}{{\cal T}}
\newcommand{\cZ}{{\cal Z}}
\newcommand{\cQ}{{\cal Q}}
\newcommand{\cB}{{\cal B}}
\newcommand{\cC}{{\cal C}}
\newcommand{\cD}{{\cal D}}
\newcommand{\cE}{{\cal E}}
\newcommand{\cF}{{\cal F}}
\newcommand{\cA}{{\cal A}}
\newcommand{\cX}{{\cal X}}
\newcommand{\IA}{\mathbb{A}}
\newcommand{\IP}{\mathbb{P}}
\newcommand{\IQ}{\mathbb{Q}}
\newcommand{\IH}{\mathbb{H}}
\newcommand{\IR}{\mathbb{R}}
\newcommand{\IC}{\mathbb{C}}
\newcommand{\IF}{\mathbb{F}}
\newcommand{\IS}{\mathbb{S}}
\newcommand{\IV}{\mathbb{V}}
\newcommand{\II}{\mathbb{I}}
\newcommand{\IT}{\mathbb{T}}
\newcommand{\IW}{\mathbb{W}}
\newcommand{\IZ}{\mathbb{Z}}
\newcommand{\re}{{\rm Re}}
\newcommand{\im}{{\rm Im}}
\newcommand{\tr}{\mathop{\rm Tr}}
\newcommand{\ch}{{\rm ch}}
\newcommand{\rk}{{\rm rk}}
\newcommand{\ext}{{\rm Ext}}
\newcommand{\bi}{\begin{itemize}}
\newcommand{\ei}{\end{itemize}}
\newcommand{\beq}{\begin{equation}}
\newcommand{\eeq}{\end{equation}}
\newcommand{\bea}{\begin{eqnarray}}
\newcommand{\eea}{\end{eqnarray}}
\newcommand{\ba}{\begin{array}}
\newcommand{\ea}{\end{array}}

\newcommand{\CN}{{\cal N}}
\newcommand{\y}{{\mathbf y}}
\newcommand{\z}{{\mathbf z}}
\newcommand{\C}{\mathbb C}\newcommand{\R}{\mathbb R}
\newcommand{\CA}{\mathbb A}
\newcommand{\CP}{\mathbb P}
\newcommand{\tmat}[1]{{\tiny \left(\begin{matrix} #1 \end{matrix}\right)}}
\newcommand{\mat}[1]{\left(\begin{matrix} #1 \end{matrix}\right)}
\newcommand{\diff}[2]{\frac{\partial #1}{\partial #2}}
\newcommand{\gen}[1]{\langle #1 \rangle}

\newcommand{\half}{\frac{1}{2}}

\newtheorem{theorem}{\bf THEOREM}
\newtheorem{proposition}{\bf PROPOSITION}
\newtheorem{observation}{\bf OBSERVATION}

\def\theequation{\thesection.\arabic{equation}}
\newcommand{\setall}{
	\setcounter{equation}{0}
}
\renewcommand{\thefootnote}{\fnsymbol{footnote}}

\begin{titlepage}
\vfill
\begin{flushright}
{\tt\normalsize KIAS-P20063}\\

\end{flushright}
\vfill

\begin{center}
{\Large\bf 5D BPS Quivers and KK Towers }

\vskip 1.5cm
Zhihao Duan\footnote{\tt xduanz@kias.re.kr},
Dongwook Ghim\footnote{\tt dghim@kias.re.kr},
and Piljin Yi\footnote{\tt piljin@kias.re.kr}
\vskip 8mm

{\it School of Physics,
Korea Institute for Advanced Study, Seoul 02455, Korea \\}

\end{center}
\vfill

\begin{abstract}
We explore BPS quivers for $D=5$ theories, compactified on a circle
and geometrically engineered over local Calabi-Yau 3-folds, for which
many of known machineries leading to (refined) indices fail due to the
fine-tuning of the superpotential. For Abelian quivers, the counting
reduces to a geometric one, but the technically challenging $L^2$
cohomology proved to be essential for sensible BPS spectra. We offer
a mathematical theorem to remedy the difficulty, but for non-Abelian
quivers, the cohomology approach itself fails because the relevant
wavefunctions are inherently gauge-theoretical. For the Cartan part
of gauge multiplets, which suffers no wall-crossing, we resort to
the D0 picture and reconstruct entire KK towers. We also perform
numerical checks using a multi-center Coulombic routine, with a simple
hypothesis on the quiver invariants, and extend this to electric BPS
states in the weak coupling chamber. We close with a comment on
known Donaldson-Thomas invariants and on how $L^2$ index might be read off from these.

\end{abstract}

\vfill
\end{titlepage}

\tableofcontents
\renewcommand{\thefootnote}{\#\arabic{footnote}}
\setcounter{footnote}{0}
\vskip 1cm

\section{$D=5$ BPS Quiver} \label{Q}

Five-dimensional gauge theories with eight supercharges \cite{Seiberg:1996bd}
can be constructed via ``compactifying" M-theory on local Calabi-Yau 3-folds
\cite{Morrison:1996xf,Intriligator:1997pq}. One can hone in for degrees of
freedom localized at the bottom of such asymptotically conical ``internal"
manifolds, while ignoring the bulk gravity. This way of realizing supersymmetric
theories via local Calabi-Yau's is broadly called geometric engineering
\cite{Klemm:1996bj,Katz:1996fh}. The gauge multiplet comes with
a single real adjoint scalar, and this renders the dynamics along the
Coulombic degrees of freedom a lot simpler and qualitatively different
from its $D=4$ counterpart, namely the Seiberg-Witten theories
\cite{Seiberg:1994rs}.  Despite such an apparent
simplification, the $D=5$ BPS spectra have been studied less vigorously.

One curious aspect of $D=5$ BPS spectra, relative to $D=4$, is the absence
of the wall-crossing.\footnote{See Ref.~\cite{Kachru:2018nck} for a recent
study of this disparity between $D=4$ and $D=5$.} One rationale behind
this is that the central charges of point-like objects are  now real,
so the well-known mechanism behind the wall-crossing is no longer viable.
What remains unclear though is exactly how this cross-over between $D=4$
and $D=5$ should be understood from the dynamics of these special objects
themselves.

It is well-known that the dynamics of BPS objects in $D=4$ is described by
quiver quantum mechanics, well-known from type IIB perspective \cite{Denef:2002ru}
but also derived directly from the Seiberg-Witten field theory near
a wall of marginal stability \cite{Lee:2011ph}.
The rank of each node is the number of the respective building
blocks, such as fundamental monopoles and dyons. The number of bifundamental
chirals between a pair of quiver nodes is determined by the Schwinger
product between the respective building blocks.
The wall-crossing is captured by the discontinuity \cite{Hori:2014tda}
of the refined Witten index \cite{Witten:1982df}
of such quantum mechanics in the parameter space of the Fayet-Iliopoulos
(FI) constants, which are in turn related to the phases of these primitive dyons.

One issue with uplifting this picture to $D=5$ is that the primitive BPS
objects, corresponding to each node of $D=4$ BPS quiver, uplift to objects with
one spatial dimension, such as BPS monopole strings; Does this mean that the
quiver quantum mechanics uplifts to $d=2$ quiver linear sigma model?

On the surface, this might sound attractive since the elliptic genus is
well-known to be safe from D-term wall-crossing \cite{Witten:1993yc}.
However, its computation relies on $T^2$, which means that $d=2$ GLSM's
suffer no such wall-crossing even if compactified on a circle; if
$D=5$ BPS objects were governed by $d=2$ linear sigma models, they
would not have experienced the wall-crossing even on $\IS^1\times \IR^{3+1}$ either.
It is by now well-known how the elliptic genus, with no wall-crossing on
the D-term parameter space, becomes piece-wise constant
only in the strict limit of $d=1$ \cite{Hori:2014tda}, while BPS
states of $D=5$ theories compactified on a circle, should experience
wall-crossing, since their central charges are still complex. So
the uplift of $D=4$ BPS quiver to $D=5$ cannot be understood as
the simple dimensional uplift of its $d=1$ quiver description.

One can also see why this naive uplift of $d=1$ GLSM to $d=2$ is a bad idea from
the simplest example of $D=4$ BPS quiver for pure $SU(2)$ Seiberg-Witten
theory, namely a Kronecker quiver with intersection number 2 \cite{Denef:2002ru}.
The two nodes represent monopoles and dyons, with charges $(0,1)$ and $(2,-1)$,
respectively. In the semi-classical picture these are a solitonic monopole
and a solitonic anti-monopole bound with a single vector meson. In
going over to $D=5$, the magnetic part uplifts to strings while the
charged vector multiplet remains as a particle. What should be noted
here is that the two objects are of opposite magnetic charge, and
thus of opposite orientations; the usual non-relativistic
approximation to extract the low energy dynamics of such solitons
no longer works.

In fact, viewing the whole situation from M-theory, the relevant objects
for magnetic sector are M5 branes wrapping 4-cycles in the Calabi-Yau,
and as such the natural theory on these string-like objects are of
$(0,4)$ supersymmetry \cite{Maldacena:1997de, Gadde:2013sca}, rather
than $(2,2)$ which would be naively suggested by the $d=1$ quiver
theories that governed $D=4$ BPS particle dynamics.

Recently, on the other hand, an interesting middle ground was offered in
Ref.~\cite{Closset:2019juk}.  The authors asked what would be the useful
low energy dynamics if one compactifies $D=5$ gauge theory on a circle
$\IS^1$ and considers particle-like states, relative to the remaining
noncompact part of the spacetime, $\IR^{3+1}$; Although the monopole
and the dyon would be still string-like, one considers only those
configurations where these strings are wrapped along $\IS^1$. These
can wiggle along the circle, but such wiggles can be attributed to
Kaluza-Klein (KK) momenta, to be treated as separate degrees of freedom.
The latter approach allows dynamics of $D=5$ BPS states to be described
yet again by some $d=1$ quiver theory, with more elementary nodes
relative to that of purely $D=4$ BPS states. The downside is that the decompactification
limit $\IS^1\rightarrow \IR^1$ requires an infinite sum over KK charges,
so the question of exactly how the wall-crossing disappears in strict $D=5$ limit
becomes a little remote.

$D=4$ BPS quiver, e.g., for pure gauge theory of rank $r$ simple gauge group,
comes with $2r$ primitive nodes typically.
The magnetic part of these $2r$ dyons is
labeled by ``simple" dual roots. Upon going up one higher dimension
with $\IS^1$, the Dynkin diagram naturally uplifts to the affine Dynkin diagram,
so we can expect two more nodes whose magnetic charges belong to
the $(r+1)$-th node of the affine Dynkin diagram. $D=5$ implies two
additional types of charges as well, namely the KK charge along $\IS^1$
and the instanton-soliton charge on $\IS^1\times \IR^3$, which should
also enter these additional nodes as well. For example, with the pure $SU(2)$
supersymmetric gauge theory, one finds a cyclic 4-node BPS quiver with
the four neighboring intersection numbers all equal to 2.

How does one figure out the quiver theory, given a $D=5$ gauge theory?
It turns out that this question has a more systematic answer than its
$D=4$ counterpart, thanks to the presence of the KK modes, i.e.,  D0 branes.
D0 probe theory for the local Calabi-Yau would be a quiver quantum mechanics,
whose  individual nodes correspond to fractional branes. These fractional
branes, M2 branes wrapping 2-cycles and M5 branes wrapping 4-cycles
times $\IS^1$, offer basic building blocks for BPS states of $D=5$
theory on $\IS^1$, so the D0 probe theory can be naturally adopted
as $D=5$ BPS quiver \cite{Closset:2019juk}.

In the context of type IIB theory, on the other hand, systematic
construction of the probe D3 theory for such local Calabi-Yau has
been pursued since late 90's, most notably by Feng, He, and Hanany \cite{Feng:2000mi}.
The construction in terms of the $d=4$ gauge theory, with four supercharges, is by now
well-established, with various techniques such as {\it Brane Tiling}
\cite{Hanany:2005ve, Franco:2005rj, Hanany:2006nm}, for toric Calabi-Yau.
As such, all one has to do is to import this technology for D0's \cite{Closset:2019juk}.

\subsection{Orbifolds}

Calabi-Yau's obtained from orbifolding $\IC^3$ by a discrete Abelian
subgroup $\Gamma$ of $SU(3)$ offer the simplest examples \cite{Douglas:1996sw, Douglas:1997de}.
For $n$ D0 probes, one starts with $n\times |\Gamma|$ many D0's and orbifolding
will lead to a quiver theory with $|\Gamma|$ many nodes, each with
gauge group $U(n)$. One can choose to assign different ranks to the nodes,
which  corresponds to adding fractional branes localized near the
orbifold point. The partial resolutions of these orbifolds are also
known to lead to other toric examples \cite{Beasley:1999uz, Feng:2000mi},
to which we will turn later in the next subsection.

One of the simplest examples of local Calabi-Yau's is local $\IP^2$ with
the asymptotic cone $\IC^3/\IZ_3$. Although this does not produce
a Seiberg-Witten gauge theory, it does offer the $D=5$ BPS quiver
in its simplest and non-trivial form. The probe theory with $n$ D0
branes starts by projecting the maximally supersymmetric $U(3n)$
Yang-Mills by $\IZ_3$ which acts on three complex coordinates by
multiplying $(w,w,w^{-2})$ with $w$ a 3rd root of unity. The
resulting quiver is a cyclic triangle quiver with three bifundamentals
for each pair of nodes.

\begin{figure}[tbp]
\begin{center}
\resizebox{0.5\hsize}{!}{
\includegraphics[height=6cm]{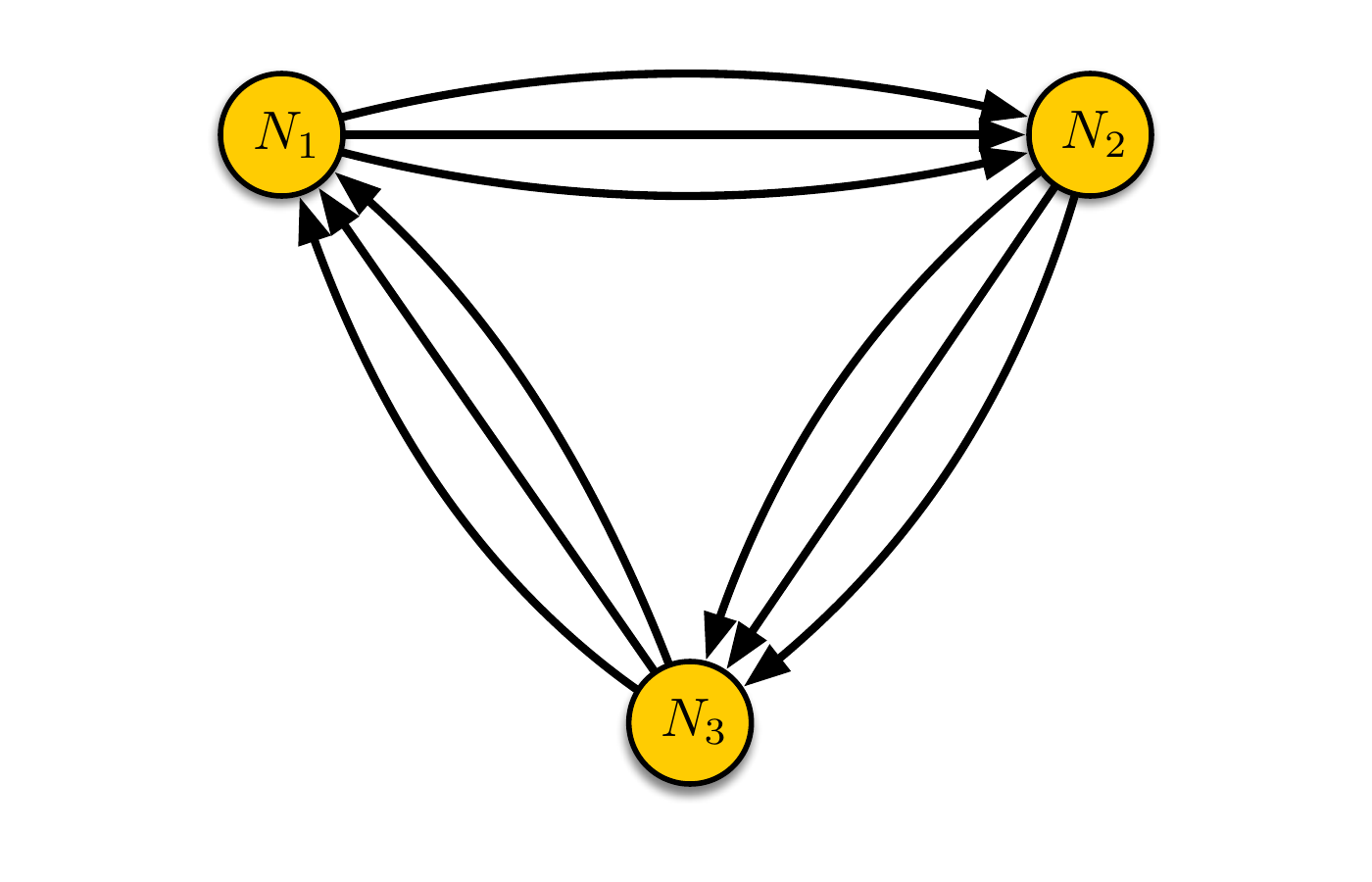}
}
\caption{\small
The quiver diagram of $\mathbb{C}^3/\mathbb{Z}_3$ orbifold theory. \label{fig:z3-quiver}}
 \end{center}
 \end{figure}

From the viewpoint of the original D0 theory prior to the orbifolding,
these bifundamental chirals can be embedded into three $U(3n)$
complex adjoints as
\bea
X=\left(\begin{array}{ccc} 0 & x_{12} & 0\\ 0 &0& x_{23} \\ x_{31} & 0 &0\end{array}\right) \ ,\;
Y=\left(\begin{array}{ccc} 0 & y_{12} & 0\\ 0 &0& y_{23} \\ y_{31} & 0 &0\end{array}\right) \ ,\;
Z=\left(\begin{array}{ccc} 0 & z_{12} & 0\\ 0 &0& z_{23} \\ z_{31} & 0 &0\end{array}\right) \,
\eea
which survive the projection
\bea
x_{ij} = w^{1+i-j}x_{ij} \ , \qquad y_{ij} = w^{1+i-j}y_{ij}\ , \qquad z_{ij} = w^{1+i-j}z_{ij}
\eea
with $w=e^{2\pi i/3}$.
The subscripts denote the pair of gauge nodes, with respect to which
these chirals are bifundamental. Note that the superpotential is not
generic but must descend from the cubic superpotential of the
maximally supersymmetric $U(3n)$ theory, i.e.,
\bea
W= \frac13{\rm tr}\left(X[Y,Z]\right)=\frac13{\rm tr}\left( x_{12}y_{23}z_{31}-x_{12}z_{23}y_{31}+\cdots\right)
\eea
where the ellipsis denotes the cyclic permutations of $x,y,z$. In fact,
this fine-tuned superpotential is a hallmark of local Calabi-Yau, and
the precise construction of $W$ has been offered through the
brane tiling machinery \cite{Hanany:2005ve, Franco:2005rj, Hanany:2006nm}.

The transition to smooth local $\IP^2$, from the singular orbifold,
is naturally described by turning on FI constants on these three $U(n)$
nodes, which corresponds to moving out into the Coulombic moduli
space of the $D=5$ theory. In a sense this is the simplest prototype
of $D=5$ BPS quivers, associated with the so-called $E_0$ theory
\cite{Morrison:1996xf}.

More elaborate examples can be found from the orbifold
$\IC^3/\IZ_p\times \IZ_p$. One projects $U(p^2n)$ theory via
\bea
x_{ij;kl} = w^{1+i-k}x_{ij;kl} \ ,\qquad y_{ij;kl} = w^{i-k}y_{ij;kl}\ ,  \qquad z_{ij;kl} = w^{-1+i-k}z_{ij;kl}\ ,
\eea
and
\bea
x_{ij;kl} = w^{j-l}x_{ij;kl} \ ,\qquad y_{ij;kl} = w^{1+j-l}y_{ij;kl} \ , \qquad z_{ij;kl} = w^{-1+j-l}z_{ij;kl}\ ,
\eea
with $w=e^{2\pi i/p}$ and the labels valued in $\IZ_p$. The surviving blocks are,
\bea
x_{ij;i+1,j}\ , \qquad y_{ij;i,j+1}\ ,\qquad z_{i+1,j+1;ij} \ ,
\eea
which are $3p^2$ bifundamentals that connect $p^2$ $U(n)$'s.

For example, the theory of $n$ D0-branes that probes $\IC^3/\IZ_2\times \IZ_2$
comes with four nodes and twelve $U(n)$ complex bifundamental chirals. Relabeling
\bea \label{orbi-node-label}
11 \rightarrow  1\ ,\quad
12 \rightarrow 2 \ , \quad
21\rightarrow 3\ ,\quad
22 \rightarrow 4 \ ,
\eea
the bifundamentals can be embedded into the  complex adjoint chirals
of $U(2^2n)$, e.g.,
\bea
X=\left(\begin{array}{cccc} 0 & 0 & x_{13} & 0 \\ 0 & 0 & 0 & x_{24}
\\ x_{31} & 0 & 0 & 0 \\ 0 & x_{42} & 0 & 0 \end{array}\right) ,\;
\eea
and similarly $Y$ and $Z$ have surviving components, respectively,
$(y_{12}, y_{21}, y_{34}, y_{43})$ and $(z_{14}, z_{23}, z_{32}, z_{41})$.

\begin{figure}[ht]
\begin{center}
\resizebox{0.5\hsize}{!}{
\includegraphics[height=6cm]{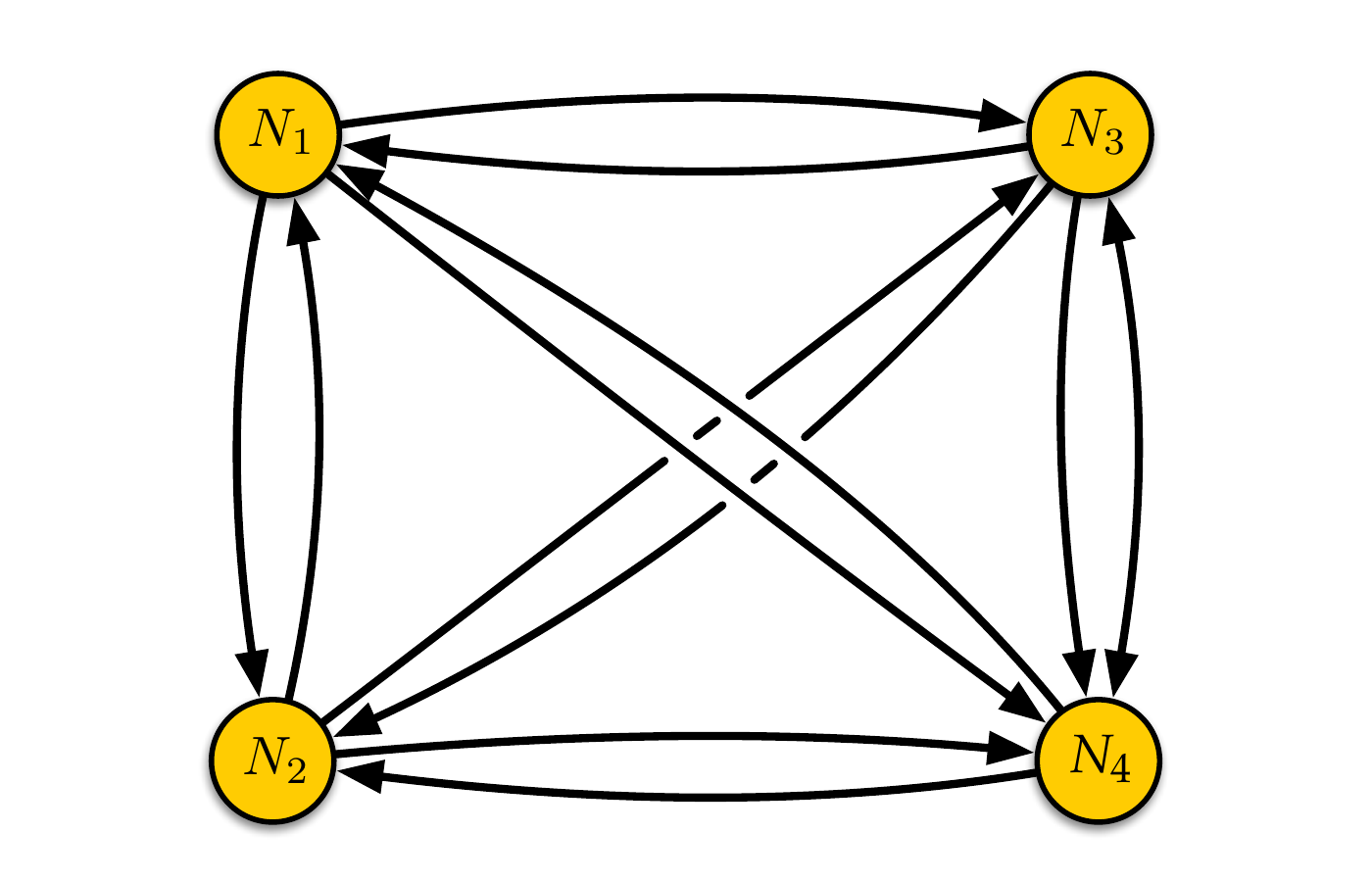}
}
\caption{\small
The quiver diagram of $\mathbb{C}^3/\mathbb{Z}_2 \times \mathbb{Z}_2$ orbifold theory. \label{fig:z2z2-quiver}}
 \end{center}
 \end{figure}

One can see that for every pair of nodes, there are two bifundamental
chirals of mutually conjugate gauge representations. Again turning
on FI parameters resolves the orbifold singularity. The
superpotential is cubic;
\bea
W= \frac{1}{2^2}{\rm tr}\left(X[Y,Z]\right)
=\frac14{\rm tr}\left( x_{13}y_{34}z_{41}-x_{13}z_{32}y_{21}+\cdots\right)\ ,
\eea
where the rest of the terms can be constructed by chasing the
subscript through the surviving block's of $X,Y,Z$, with
the sign determined by the parity of the permutation of $x,y,z$.
Extension of this to $\IZ_p\times\IZ_p$ is straightforward.
The relabeling of $ij$ for $p>2$,
\bea \label{orbi-node}
ij \rightarrow (i-1)p+j
\eea
embeds three sets of the $p^2$ bifundamental chirals into
three $U(p^2n)$ complex adjoint chirals, $X,Y,Z$, respectively,
and $W={\rm tr}\left(X[Y,Z]\right)/p^2 $ contains
total $2p^2$ monomials, half of which have $1/p^2$ and the
other half $-1/p^2$ as the coefficient.

A special subsector of this orbifold dynamics emerges when we
force all the surviving blocks to take values of three common
$n\times n$ matrices, $x_*, y_*, z_*$. The superpotential will
collapse to a single cubic commutator potential,
\bea
W\rightarrow {\rm tr}\left(x_* [y_* , z_*]\right)\ ,
\eea
and the three chirals are each complex adjoint relative to the
``diagonal" $U(n)_*$ which rotates $U(n)$'s sitting at nodes all
simultaneously. This $U(n)_*$ together with $x_*,y_*,z_*$
defines maximally supersymmetric
$U(n)$ Yang-Mills theory. Furthermore, the overall $U(1)$
and trace parts of these three complex chirals decouple from
the rest of $U(n)_*$ theory, and leave behind a maximally
supersymmetric $SU(n)$ Yang-Mills. Let us call the latter $SU(n)_*$ theory.

This $SU(n)_*$ theory has a clear interpretation as the local
and relative dynamics  of $n$ D0-branes. When D0's are clustered
near each other and sitting at a generic point of the Calabi-Yau,
they will see the spacetime locally as $\IR^{10}$ unless they
are at the top of a singular point; So long as the mutual distances
between D0's are sufficiently small, the dynamics among them would
be controlled by the flat space D0 theory. This local dynamics $SU(n)_*$
will prove to be a key to obtaining the Kaluza-Klein towers due to $\IS^1$.

\subsection{Partial Resolution of Orbifolds} \label{ex-quiver}

More examples of local toric Calabi-Yau's and the accompanying $D=5$ BPS
quiver can be found by starting with an orbifold and resolving the singularity
partially \cite{Beasley:1999uz, Feng:2000mi}.
This procedure is called ``Higgsing" since
the resolution involves turning on FI constants partially such that
some bifundamental chirals acquire vacuum expectation values.

Among various Calabi-Yau 3-folds which geometrically engineer $D=5$ gauge theory, we introduce two classes of geometry, known as the $Y^{p,q}$ and $X^{p,q}$ families \cite{Gauntlett:2004yd,Benvenuti:2004dy, Martelli:2005wy, Hanany:2005hq},
named after the Sasaki-Einstein 5-manifold which occupies their angular direction.
We first discuss the former, of which $p=2, q=0$ case ($Y^{2,0}$) is particularly known as local $\mathbb{F}_0$ surface, and the latter will be discussed shortly.

\subsubsection*{$Y^{p,q}$ family and local $\mathbb{F}_0$ geometry}

\begin{figure}[htbp]
\begin{center}
\resizebox{0.88\hsize}{!}{
\includegraphics[height=8cm]{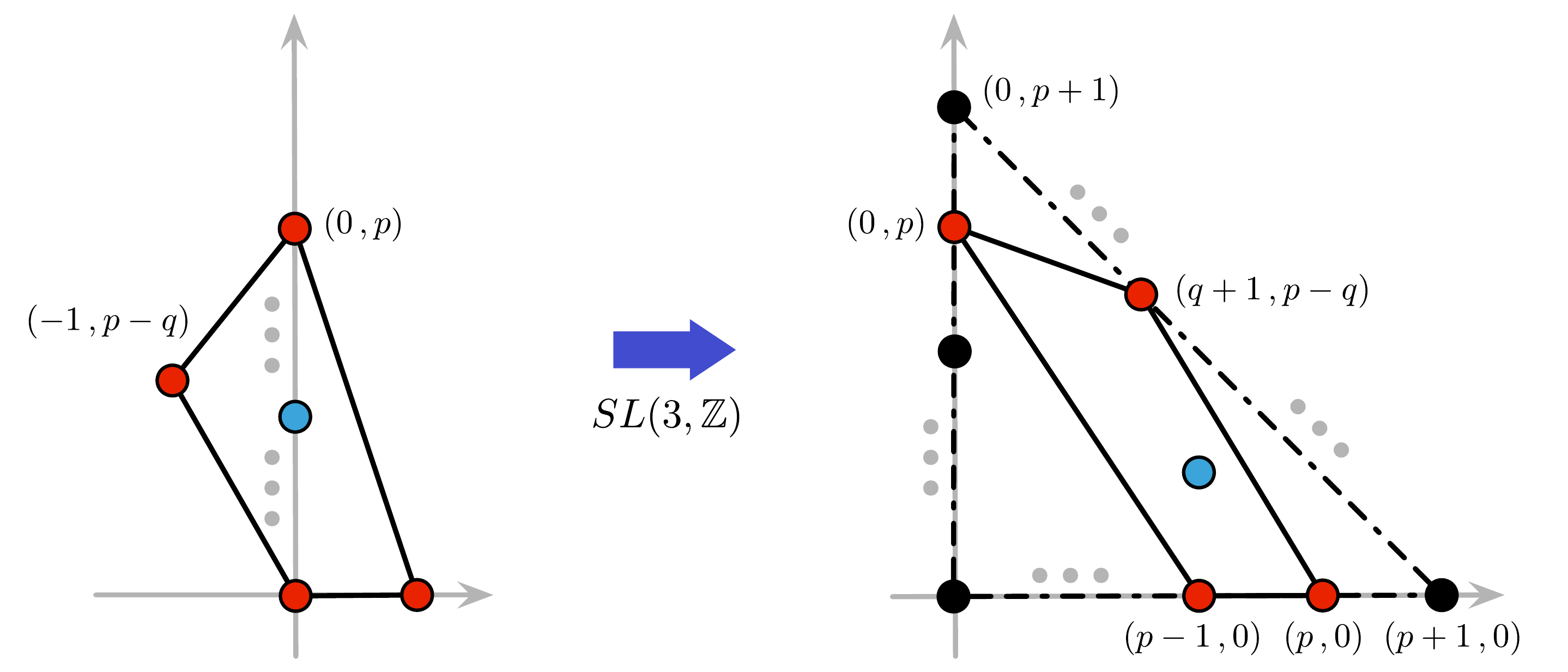}
}
\caption{\small
An embedding of $Y^{p,q}$ toric diagram in the toric diagram of $\mathbb{C}^3 / \mathbb{Z}_{p+1} \times \mathbb{Z}_{p+1}$ orbifold theory. The linearly dotted gray points imply several internal vertices which are omitted for brevity. \label{fig:ypq-toric-embed}}
\end{center}
\end{figure}

Given two non-negative integers $p$ and $q$ such that $p > q \geq 0\,,$
a Calabi-Yau 3-fold $Y^{p,q}$ describes a fibration of the ALE space
of $A_{p-1}$ type over $\mathbb{P}^1$.
It geometrically engineers $D=5$ $\cN=1$ $SU(p)_q$ gauge theory \cite{Hanany:2005hq}.
The toric diagram of $Y^{p,q}$ 3-fold is given by the four external vertices\footnote{
If necessary, we triangulate the diagram by connecting
$v_1$ and $v_4$ to all internal points along the $y$-axis, together with vertical segments.
This resolves the singularity hence the geometry becomes smooth.
The other triangulations are related by flop transitions which do not affect our discussion
in the paper, except for the detailed dictionary between internal cycles in resolved CY's and fractional branes in Appendix \ref{map}.},
\begin{align} \label{ypq-toric-coord}
v_1 = (1,0), \quad v_2 = (0,0), \quad v_3 = (0,p), \quad v_4 = (-1, p-q)\,.
\end{align}
Figure \ref{fig:ypq-toric-embed} shows how to embed the toric diagram \eqref{ypq-toric-coord} in a larger triangular toric diagram,
corresponding to $\mathbb{Z}_{p+1} \times \mathbb{Z}_{p+1}$ orbifold
after an $SL(3,Z)$ transformation.
This suggests a way of taking partial resolution\footnote{Note that we need to freeze $(p^2+1)$-many nodes before reaching to
$Y^{p,q}$ quiver. It remains to be seen whether there would be
a more efficient embedding, i.e. toric diagram of orbifold theory
with smaller rank, to cast $Y^{p,q}$ toric diagram.} of orbifold probe theory
in order to obtain the $D=5$ BPS quiver of $SU(p)_q$ gauge theory.

The simplest example in this class would be local $\mathbb{F}_0$ surface, or $Y^{2,0}$, which engineers $D=5$ $SU(2)$ pure gauge theory.
The detailed procedure of Higgsing toward $F_0$ is given as follows.
Relabeling nodes of $\mathbb{C}^3/ \mathbb{Z}_3 \times \mathbb{Z}_3 $ orbifold theory following \eqref{orbi-node},
we have the superpotential
\begin{align} \label{z3z3-w}
    W = \frac{1}{3^2} \text{tr} \left( x_{14} y_{45} z_{51} - x_{14} z_{43} y_{31} + \cdots + x_{93} y_{31} z_{19} - x_{93} z_{38} y_{89} \right) \,.
\end{align}
A vev $v$ assigned to $z_{51}$ merges node $1$ and node $5$ of the orbifold theory as their relative $U(n)$ is frozen.
By integrating out massive fields $x_{14} \,, x_{25} \,, y_{12} \,, y_{45}$,
we have superpotential
\begin{align}
\begin{split}
    \frac{1}{9} \text{tr} & \left( v^{-1} y_{16} z_{62} z_{27} x_{71}  - v^{-1} x_{18} z_{84} z_{43} y_{31}  + x_{36} y_{64} z_{43} - x_{36} z_{62} y_{23} \right. \\
    &  + x_{47} y_{78} z_{84} - x_{47} z_{76} y_{64} + x_{18} y_{89} z_{91} + x_{69} y_{97} z_{76}  - x_{69} z_{91} y_{16}  \\
    & \left.- x_{71} z_{19} y_{97} + x_{82} y_{23} z_{38} - x_{82} z_{27} y_{78} + x_{93} y_{31} z_{19} - x_{93} z_{38} y_{89} \right)
\end{split}
\end{align}
where the label $5$ is now renamed as $1$.
By repeating similar exercise with four more FI constants turned on so that $x_{93} \,, y_{23} \,, y_{64} \,, x_{18}$ chiral fields are Higgsed one by one,
we obtain the following superpotential
\begin{align} \label{f0-w}
   W_{\mathbb{F}_0} = \text{tr} \left( x_{23} y_{34} y_{41} z_{12} -x_{23} z_{34} y_{41} y_{12} + x_{41} y_{12} z_{23} z_{34} - x_{41} z_{12} z_{23} y_{34}  \right) \,.
\end{align}
\begin{figure}[tp!]
\begin{center}
\resizebox{0.5\hsize}{!}{
\includegraphics[height=6cm]{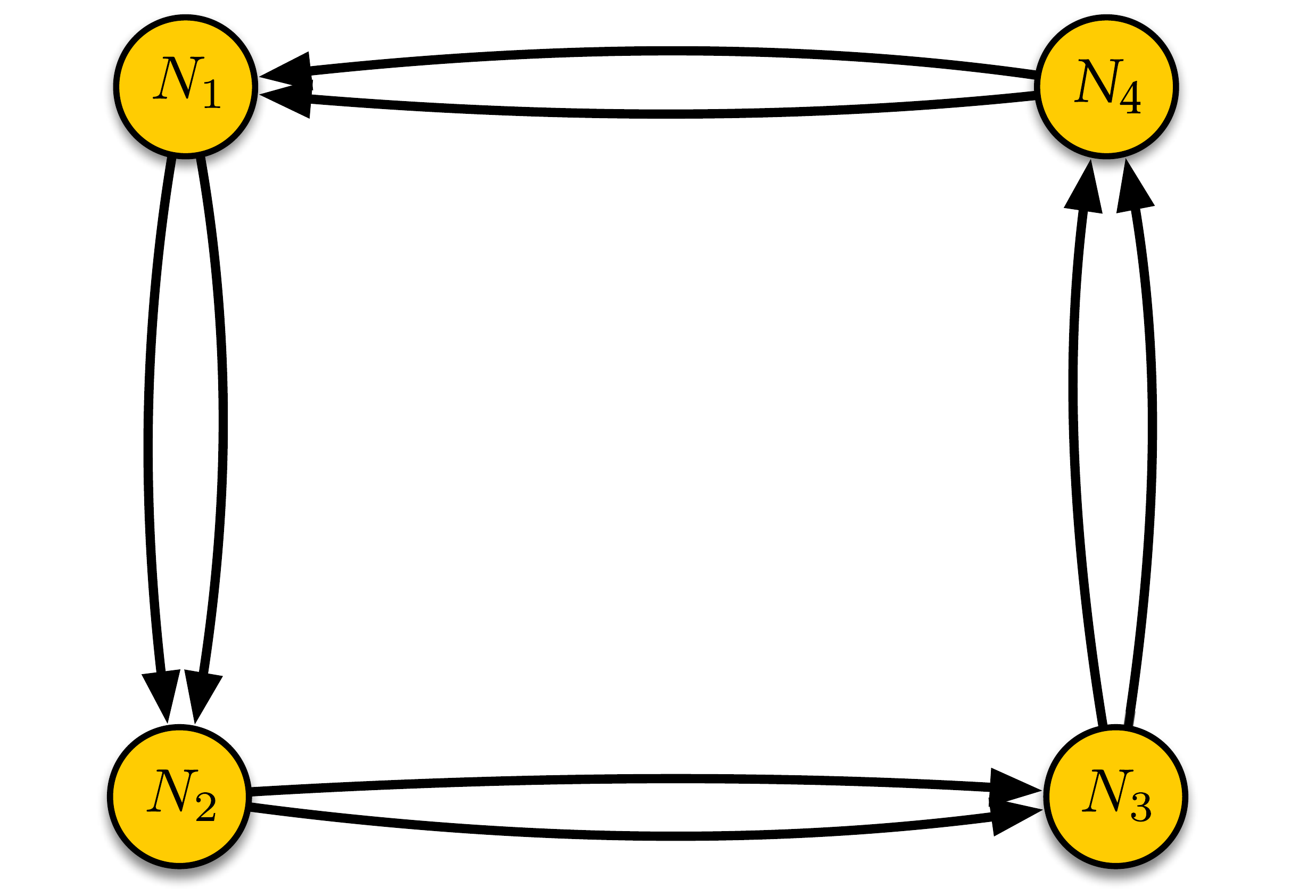}
}
\caption{\small
The quiver diagram of local $\mathbb{F}_0$ theory \label{fig:f0-quiver}}
 \end{center}
\end{figure}
Note we rescaled the chiral fields to avoid clutter then relabeled the nodes as follows,
\begin{align}
    5 \,, 8 \rightarrow 1 \,, \quad 6 \rightarrow 2 \,, \quad  3 \,, 9 \rightarrow 3 \,, \quad 7 \rightarrow 4 \,.
\end{align}
The quiver diagram of local $\mathbb{F}_0$ theory is drawn in Figure \ref{fig:f0-quiver}.

\begin{figure}[ht!]
\begin{center}
\resizebox{0.8\hsize}{!}{
\includegraphics[height=7cm]{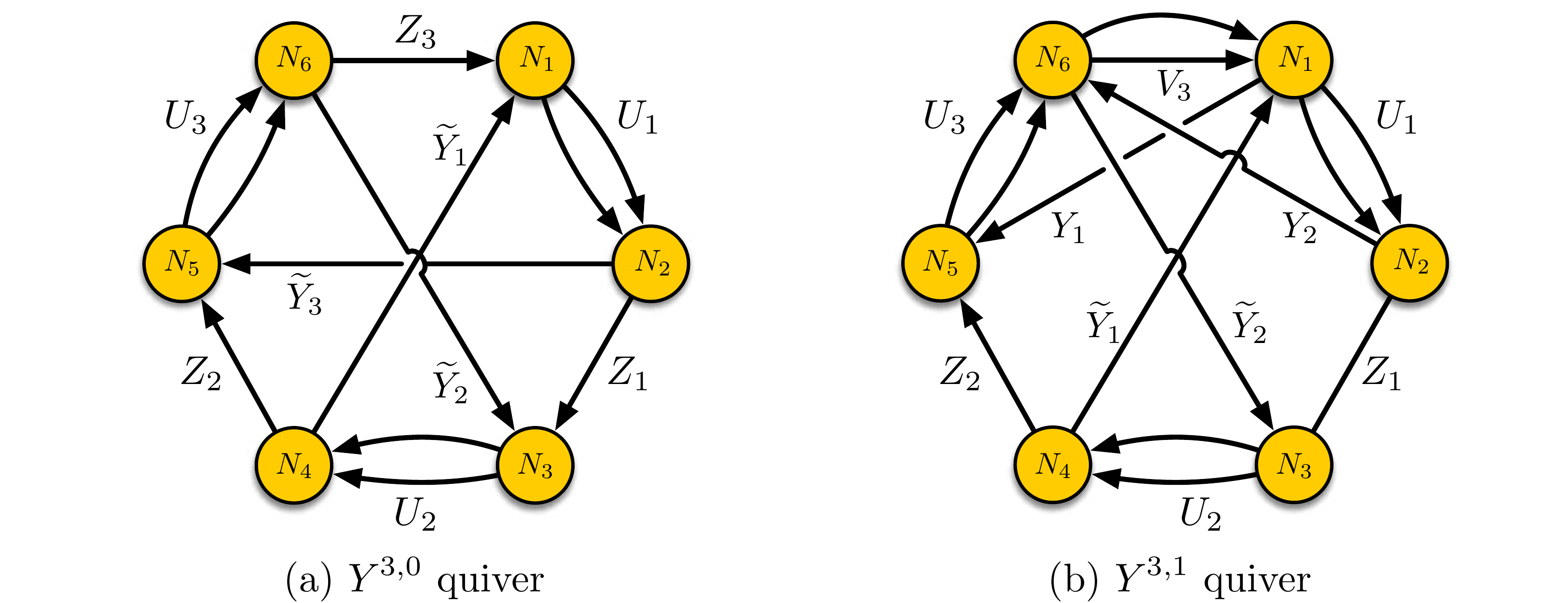}
}
\caption{\small
Quiver diagrams of $Y^{p,q}$ theory (a) with $p=3$, $q=0$ and (b) with $p=3$ , $q=1$. The chiral fields are labelled by the convention in \cite{Closset:2019juk}. \label{fig:yp=3-quiver}}
\end{center}
\end{figure}

A $D=5$ BPS quiver of higher rank gauge theory is obtained by starting with
larger orbifold theories and by taking a similar Higgsing procedure, which
truncates the corners of toric diagram according to Figure \ref{fig:ypq-toric-embed}.
For example, the BPS quivers of $SU(3)$ gauge theory are obtained by
assigning vev to ten chiral fields in the orbifold probe quiver
$\mathbb{C}^3 / \mathbb{Z}_4 \times \mathbb{Z}_4$.
Figure \ref{fig:yp=3-quiver} shows resulting $D=5$ BPS quivers for $SU(3)_0$ and $SU(3)_1$ gauge theory, respectively.

\subsubsection*{$X^{p,q}$ family and local $dP_2$ surface}

\begin{figure}[ht]
\begin{center}
\resizebox{0.88\hsize}{!}{
\includegraphics[height=8cm]{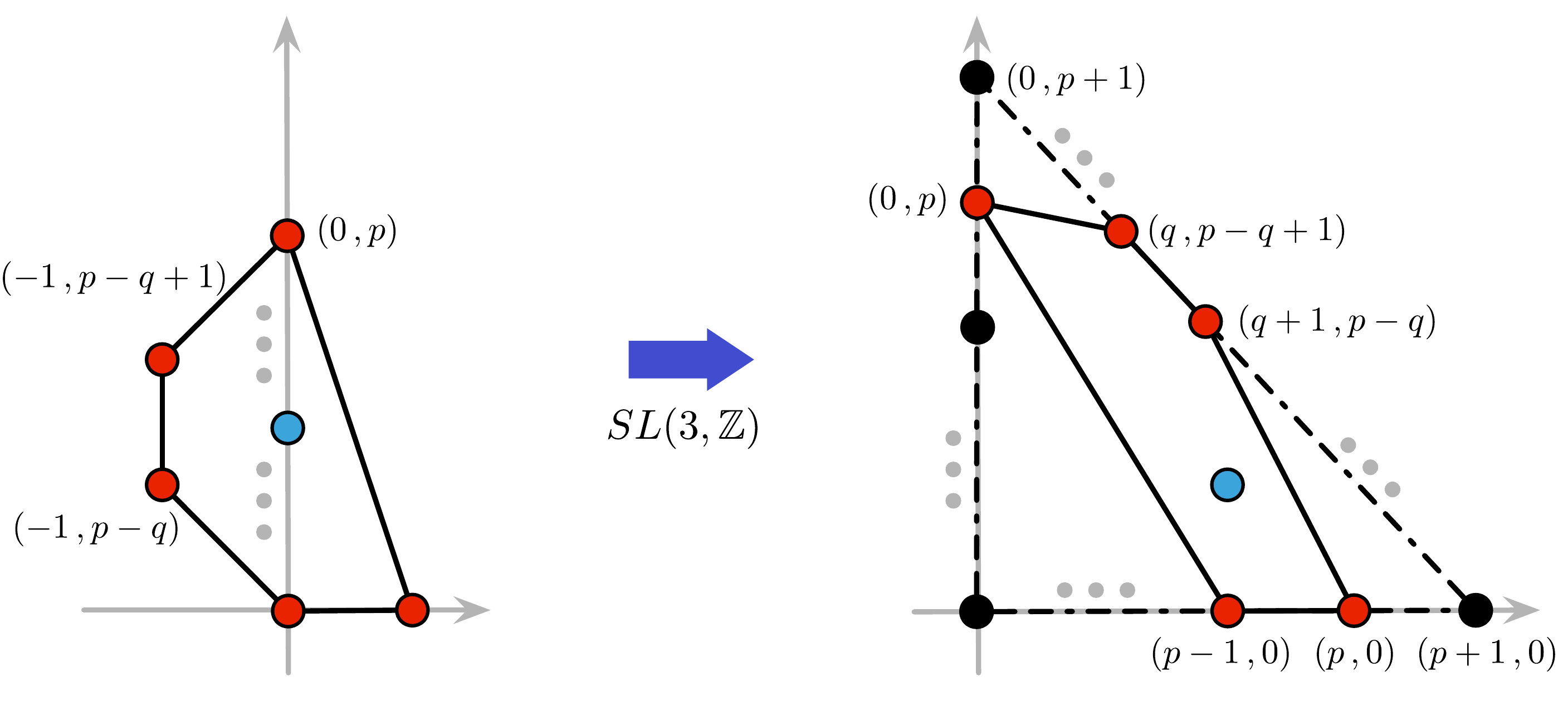}
}
\caption{\small
An embedding of $X^{p,q}$ toric diagram in the toric diagram of $\mathbb{C}^3 / \mathbb{Z}_{p+1} \times \mathbb{Z}_{p+1}$ orbifold theory. The linearly dotted gray points imply several internal vertices which are omitted for brevity. \label{fig:xpq-toric-embed}}
\end{center}
\end{figure}

Another class of toric Calabi-Yau 3-folds of interest is $X^{p,q}$ family, for two integers $p$ and $q$ such that $p>q \geq 1$. Its toric diagram has five external vertices
\begin{align} \label{xpq-toric-coord}
v_1 = (1,0), \quad v_2 = (0,0), \quad v_3 = (0,p), \quad v_4 = (-1, p-q), \quad v_5 = ( -1, p-q+1) \,.
\end{align}
An embedding of \eqref{xpq-toric-coord} to $\mathbb{Z}_{p+1} \times \mathbb{Z}_{p+1}$ orbifold's toric diagram is sketched in Figure \ref{fig:xpq-toric-embed}, which suggests how to obtain $D=5$ BPS quiver of $SU(p)_q$ gauge theory with a fundamental hypermultiplet,
starting from the orbifold probe quiver via Higgsing.

\begin{figure}[ht]
\begin{center}
\resizebox{0.5\hsize}{!}{
\includegraphics[height=6cm]{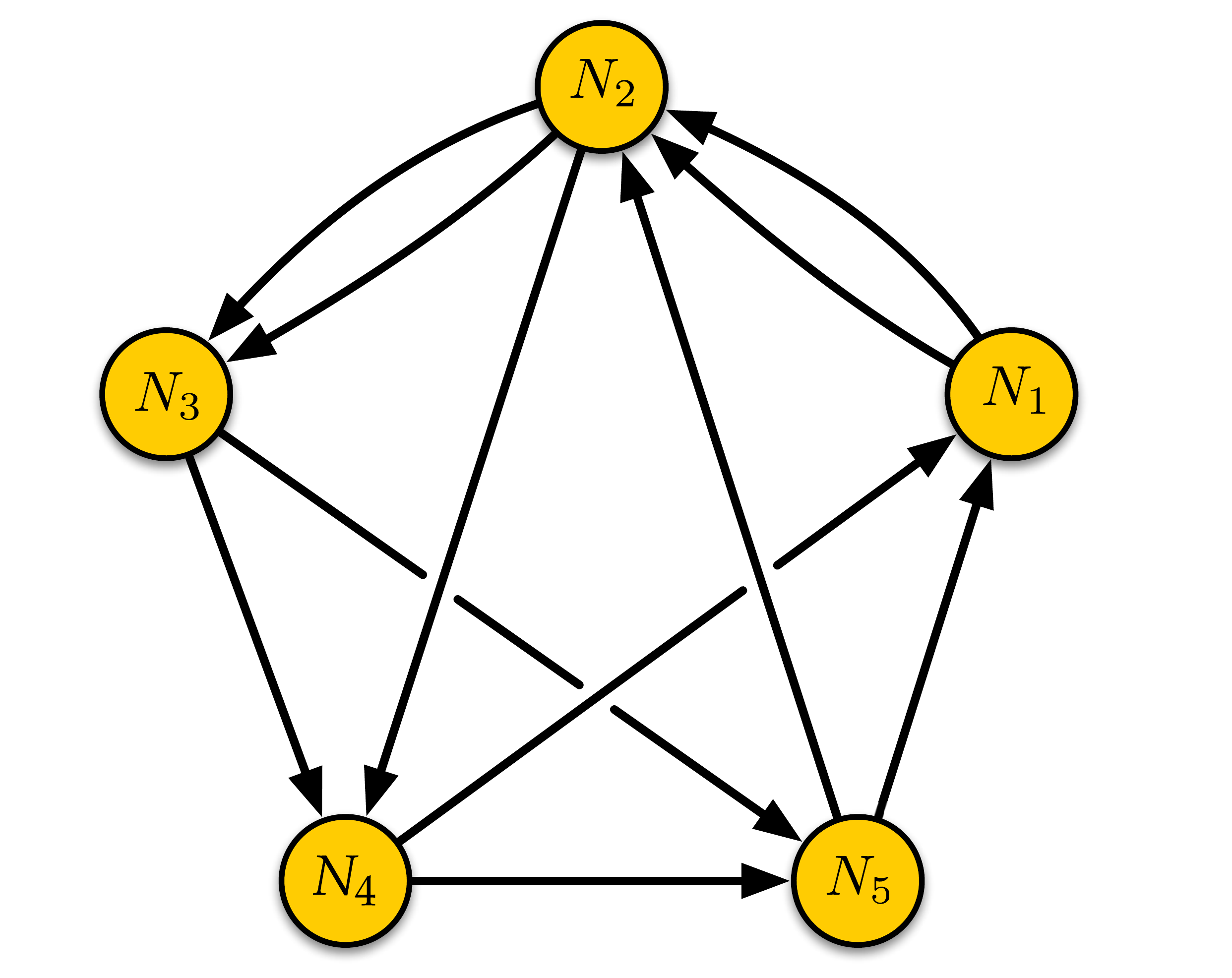}
}
\caption{\small
The quiver diagram of local $dP_2$ theory.\label{fig:dp2-quiver}}
 \end{center}
 \end{figure}

The simplest example in this family is $X^{2,1}$, also known as the local $dP_2$ geometry. Its BPS quiver can be explicitly worked out, as shown in Figure \ref{fig:dp2-quiver}.
This geometry turns out to engineer $D=5$ $SU(2)$ gauge theory with a fundamental hypermultiplet \cite{Hanany:2005hq}.
From the superpotential of $\mathbb{Z}_3 \times \mathbb{Z}_3$ orbifold theory
\begin{align} \nn
    W = \frac{1}{3^2} \text{tr} \left( x_{14} y_{45} z_{51} - x_{14} z_{43} y_{31} + \cdots + x_{93} y_{31} z_{19} - x_{93} z_{38} y_{89} \right) \,,
\end{align}
non-zero FI constants can lead vev assignment to $z_{51} \,, x_{93} \,, y_{23} \,, y_{64}$,
which makes neighboring fields in the quiver massive.
As we integrate out those massive, the superpotential becomes
\begin{align}
\begin{split}
    W_{dP_2} = \text{tr} & \left (  x_{34} y_{41} z_{12} z_{23}  + x_{12} y_{23} y_{35} z_{51}   - x_{45} z_{51} z_{12} y_{23} x_{34} \right. \\
        & \left.  + x_{45} x_{52} z_{24} - x_{12} z_{24} y_{41} - x_{52} z_{23} y_{35}  \right) \,,
\end{split}
\end{align}
where we relabeled the nodes in the quiver as follows,
\begin{align}
        6 \rightarrow 1 \,, \quad 3 \,, 9 \rightarrow 2 \,, \quad
        7 \rightarrow 3 \,, \quad 5 \rightarrow 4 \,, \quad 8 \rightarrow 5.
\end{align}
Note that if we further Higgs $x_{45}$ fields, we end up with local $\mathbb{F}_0$ theory
superpotential in \eqref{f0-w} and quiver in Figure \ref{fig:f0-quiver} as the node $4$ and $5$ merge.

\section{What to Count?} \label{problem}

The problem we wish to address is the counting of BPS states which
are particle-like
with respect to the noncompact part of the spacetime, i.e., $\IR^{3+1}$.
Given the quivers of Section \ref{Q}, it comes down to computing various
topological invariants of such quiver dynamics such as refined
Witten index, or cohomology if the dynamics reduces to a geometric
one in the low energy limit. In practice, the former reduces to
an exact path integral \cite{Hori:2014tda} or to a heat kernel computation
of twisted partition function, while the latter can be
sometimes computed explicitly if the geometry is simple
enough, e.g., toric.

While either of such methods would work perfectly well when the
dynamics is fully gapped, there is a general problem when
the dynamics includes asymptotically flat directions. In particular,
the problem arises invariably if the low energy theory involves
an infinite volume of target manifold. All of the current examples
involve local Calabi-Yau as part of the target, so such subtleties
occur generically.

This is further compounded by the fine-tuned superpotential
that is generically needed for local Calabi-Yau's.  The well-known
machinery for exact path integrals, often broadly called
``localization," does not seemingly depend on details of
the superpotential, but only with the hidden assumption of
the genericity. Once we begin to impose non-Abelian
global symmetries on the superpotential that cannot be entirely
captured by flavor chemical potentials, the result of the
localization computation often deviates significantly
from the desired counting \cite{Lee:2016dbm}. While the latter
problem is also obliquely related to the noncompact nature
of the target, it should be considered  as a separate issue,
and one can see by explicit exercises that the resulting JK
residue computation \cite{Hori:2014tda} becomes rather
ineffective even for  the simplest examples.

Coming back to the specific counting problem at hand, note that the
particle-like nature of BPS states means that the wavefunction of the states must be restricted
to the $L^2$ class, i.e., square-normalizable with respect to the internal
directions, while along $\IR^{3+1}$ the state would propagate freely
as plane wave. Even with more conventional and geometric approaches,
such $L^2$ boundary condition is difficult to impose systematically
for index problems.
One well-known exception would be an asymptotically cylindrical
non-linear sigma model (NLSM) where $L^2$ condition translates to the
Atiyah-Patodi-Singer (APS) boundary condition \cite{Atiyah:1973aps}.

As such, we will encounter two types of topological invariants.
One class, which in this note would encompass many different types,
computed by relatively straightforward existing routines,
will be denoted as $\Omega$. It could be a result of the localization
path integral or of the heat kernel computation; When the problem
reduces to an entirely geometric one, such as via cohomology,
we will also use $\Omega$ to denote index-like objects of
singular homology or (compact) de Rham cohomology as well.
The other, denoted as $\cI$, would be the desired (refined)
Witten index that counts physically relevant $L^2$
BPS states only.

The two objects are fundamentally different,
despite that physics literatures tend to refer both as ``index."
Only if the dynamics is compact and the Hilbert space
is fully gapped, $\Omega=\cI$ is guaranteed.
Much of what we explore in this note would be about how to
recognize the difference, and, sometimes, how to extract $\cI$ from $\Omega$
for these quiver theories.

In the path integral computation of the refined index, say, for
those examples with fine-tuned superpotentials, one can seemingly
evade such issues by turning on flavor chemical potentials
associated with isometries. This essentially turns on mass terms
and creates a gap. While the resulting partition function $\Omega$
has its own physical interpretation and is often very useful, e.g.,
for a check of dualities, such twisted partition functions sometimes
give misleading results when it comes to the $L^2$ BPS state counting \cite{Lee:2016dbm}.

An almost trivial, yet illustrative example is
a NLSM with target $\IC^1$, say, realized
in terms of a single massless chiral multiplet. The theory
has a global symmetry that rotates $\IC^1$ by a phase. Assigning
a fugacity $x$ to this symmetry, likewise ${\bf y}$ for an $R$-symmetry,
and vanishing R-charge to the single chiral multiplet to prohibit
any superpotential, one finds the following twisted partition
function from the localization of the path integral \cite{Hori:2014tda},
\bea
\Omega_{\IC^1} = \frac{{\bf y}x^{-1/2}-{\bf y}^{-1}x^{1/2}}{x^{1/2}-x^{-1/2}}\,.
\eea
Turning off the flavor chemical potential, $x\rightarrow 1$,  is
impossible due to the pole,  symptomatic of the noncompact
dynamics.

Expanding in $x$ or $1/x$  might be the next tempting step,
since $x$ enters the definition of $\Omega$ in the form of $x^F$
where $F$ is the flavor charge. This will at least give $F$-graded
sectors, and the zeroth power of $x$ may have some chance
of capturing the physics prior to the massive deformation due to $x$.
However, the two expansions yield
mutually inconsistent results at the invariant sector
\bea \label{C1}
\Omega_{\IC^1}=\left\{\begin{array}{l} -{\bf y}^{-1}+O(1/x) \,, \\ -{\bf y}+ O(x) \,. \\ \end{array} \right.
\eea
A general relation for fully gapped geometric theories says \cite{Lee:2012naa}
\bea
\cI=(-{\bf y})^d\chi_{y=-\bf y^2}\, ,
\eea
where $d$ is the complex dimension
of the compact target and $\chi_y$ is the Hirzebruch genus.
The R-charge fugacity ${\bf y}$
grades the Hodge diamond diagonally, which should extend to
any geometric theories, compact or not. The above
expansion of $\Omega$ suggests that the content of the rotationally
invariant BPS state corresponds to either one-dimensional $H^2$ or
one-dimensional $H^0$, with all other cohomology empty. This follows
from (\ref{C1}), combined with the fact that $\IC^1$ is algebraic
such that the Hodge diamond is populated by the vertical middle only.

Interestingly, these two conflicting results appear to reflect,
respectively, de Rham cohomology $H^\star$ and de Rham cohomology
with compact support $H^\star_c$,
\bea
&&{\rm dim}\,H^0(\IC^1)= {\rm dim}\,H^2_c(\IC^1)=1 \ ;\cr\cr
&& {\rm dim}\,H^k(\IC^1)= {\rm dim}\,H^{2-k}_c(\IC^1)=0\ , \quad k=1,2\,.
\eea
As we noted above, however, the physically relevant cohomology
would be $\cH^\star_{L^2}$, whereby one is supposed to count $L^2$
harmonic forms on the target, which is clearly absent for $\IC^1$,
\bea
\cH^\star_{L^2}(\IC^1)= 0  \quad\rightarrow\quad \cI=0\,.
\eea
This trivial example teaches us that neither the localization
computation nor the standard (compact) de Rham cohomology
computation should be trusted.

A more informative example can be found from an Abelian
GLSM with $N$ chirals of charge $+1$ and $K$ chirals of charge $-1$. Suppose
$N>K$ for simplicity and the sign of the FI constant such that this
GLSM flows down to the $\mathcal{O}(-1)^K$ bundle over $\IP^{N-1}$. Keeping a
flavor symmetry that rotates all chirals simultaneously and forbidding
superpotential by assigning zero $R$-charges to all chirals, we expand
with respect to its fugacity $x$ and find
\bea
\Omega=\left\{\begin{array}{l}
(-1)^{N-K-1}\left({\bf y}^{1-K-N}+\cdots +{\bf y}^{N-K-1}\right)+O(1/x) \,, \\ \\
(-1)^{N-K-1}\left({\bf y}^{1+K-N}+\cdots +{\bf y}^{N+K-1}\right)+O(x)\,. \\\end{array}\right.
\eea
The flavor-neutral part, or the $x$-independent part, captures
again precisely $H^\star$ and $H^\star_c$, respectively, of the bundle.
The latter statement can be seen easily from the fact that $H^\star$ is homotopy
invariant, so $H^\star$ of the bundle equals $H^\star(\IP^{N-1})$, and
also from the general fact that $H^\star_c$ is a Poincar\'e dual of
$H^\star$.

What is $\cH_{L^2}^\star$ of this bundle? As we will see in Sec. \ref{secTHM}, there
is a mathematical theorem which asserts that, under some favorable
circumstances, $\cH_{L^2}^p$ equals $H^p$ for $p>d$ where $d=N+K-1$ is the
complex dimension. Also, the $L^2$ cohomology should come
with natural pairing between $\cH_{L^2}^p$ and $\cH_{L^2}^{2d-p}$, forming
Poincar\'e-dual pairs. This leaves behind only $\cH^d_{L^2}$ undetermined.
If $d$ happens to be odd, we can say more, since  only even cohomologies are non-trivial,
\bea
&&{\rm dim}\cH_{L^2}^{2K}= {\rm dim}\cH_{L^2}^{2K+2}=\cdots
={\rm dim}\cH_{L^2}^{2N-2}=1 \ ;\cr\cr
&& {\rm dim}\cH_{L^2}^p=0\  ,\quad \hbox{for all other } p \,.
\eea
These $L^2$ states can be compactly summarized via a truly
enumerative index
\bea
\cI=(-1)^{N-K-1}\left({\bf y}^{1+K-N}+\cdots +{\bf y}^{N-K-1}\right)\,,
\eea
which is the rescaled Hirzebruch genus of the $L^2$ cohomology.
One curious fact is that this result could have been obtained
if we blindly took the common part of the two expansions of $\Omega$ in
$x$ and $1/x$ as above \cite{Lee:2016dbm}, or equivalently the
common part of $H^\star$ and $H^\star_c$. The latter is effectively
the content of the theorem (\ref{L_2thm}) below.

As these examples illustrate well, distinctive features
of $\cI$, as opposed to generic $\Omega$, are that it is
integral, finite, and a symmetric Laurent polynomial in the $R$-charge fugacity
${\bf y}$. Note that, for BPS quivers in particular, the
power of ${\bf y}$ is related to the spin and
$SU(2)_R$-isospin content of the BPS spectra \cite{Lee:2012naa}, so
the invariance under  ${\bf y}\rightarrow {\bf y}^{-1}$ is also a
consequence of the spacetime symmetries on $\IR^{3+1}$.

\section{Abelian BPS Quivers} \label{D0}

The main message above was how one must not be hasty and
not resort to existing routines for supersymmetric partition
functions or to (co-)homology of the internal manifold. When the
latter is noncompact, the partition function will generically
give misleading numbers while one must be also wary of exactly
which cohomology is relevant for the problem at hand.
As such, for some questions, the only reliable answer can be obtained
from the direct cohomology counting,  assuming that the distinction
between $H^\star$, $H^\star_c$, and $\cH_{L^2}^\star$ is carefully kept
track of. For the toric case, on the other hand, there are systematic
tools, available in Ref.~\cite{sage} for example, that recover the
homology $H_\star$ given the toric data, from
which $H^\star$ can be inferred. We will presently turn to how one
might further extract $\cH_{L^2}^\star$ from such data.

We should warn the readers that even this is not going to be
effective when the quiver turns non-Abelian; the low energy
geometric limit of the latter often misses the relevant
ground states of the quiver gauge theory as will be outlined
in Section~\ref{nD0}.

\subsection{$L^2$ Cohomology from Homology}\label{secTHM}

Let us start by reviewing a theorem in \cite{Hausel:2002xg}
which discussed relations between $L^2$ cohomology and
following types of cohomology given a noncompact manifold $\cM$
with scattering metric $g$ assigned;
\begin{itemize}
    \item de Rham cohomology $H^\star (\cM)$
    \item relative cohomology with respect to the boundary $H^\star (\cM, \partial \cM)$
    \item de Rham cohomology with compact support $H^\star_c (\cM) $
\end{itemize}
A metric $g$ on a manifold $\cM$ is called the \textit{scattering} metric when $g$ satisfies the following asymptotic behavior,
\begin{align} \label{g-scatt}
g \, \rightarrow \,  \frac{dx^2}{x^4} + \frac{h}{x^2} \,,
\end{align}
where $x$ is a boundary defining function, i.e. $x$ vanishes but $dx \neq 0 $ on $\partial \cM$ and $h$ is a smooth metric on $\partial \cM$.
Note that if we set $x=1/r$, the scattering metric in \eqref{g-scatt} becomes
\begin{align}
    g \rightarrow dr^2 + r^2 h \qquad \text{with} \quad r \rightarrow \infty \,,
\end{align}
which reproduces a familiar form of metric on a conical geometry.

For a manifold $\cM$ with scattering metric $g$, there exist natural isomorphisms,
\begin{align}\label{L_2thm}
    \cH_{L^2}^p (\cM) \rightarrow \begin{cases}
    H^p (\cM \,, \partial \cM) \,, & \text{if $p < m/2 $} \,, \\
    \text{Im} ( H^p (\cM \,, \partial \cM) \rightarrow H^p (\cM) ) \,, & \text{if $p=m/2$} \,, \\
    H^p (\cM) \,, & \text{if $p > m/2$} \, ,
  \end{cases}
\end{align}
with $m={\rm dim}_\mathbb{R} \cM$, where $\cH^p (\cM \,, \partial \cM)$
denotes the relative cohomology of $\cM$ with respect to its boundary.
For a complex manifold, such as Calabi-Yau $d$-folds, with $m=2d$,
this means
that $\cH^p_{L^2}(\cM)$ equals $H^p(\cM)$ for $p>d$, determining
the upper half of Betti numbers. This in turn determines
the lower half as well since there is a natural pairing,
\bea
\int_\cM \omega_{(k)}\wedge \omega_{(2d-k)}\ ,
\eea
which leaves only the middle cohomology $\cH_{L^2}^d(\cM)$ to be
further considered. For our problem of local Calabi-Yau 3-fold,
which are toric and thus algebraic, homology $H_{3}(\cM)$ is empty, so both $H^{3}(\cM)$
and $\cH_{L^2}^3(\cM)$ are also empty.\footnote{Although the
Calabi-Yau property implies a covariantly constant holomorphic
3-form, it does not generate the ordinary de Rham $H^3$
for local Calabi-Yau's which are toric. This can be glimpsed at,
if somewhat trivially, with the example of $\IC^3$. The
holomorphic 3-form, $\Omega^{3.0}=dz_1\wedge dz_2\wedge dz_3$
can be also written as $\Omega^{3,0}=d(z_1\,dz_2\wedge dz_3)$
and hence is exact in the absence of an asymptotic boundary
condition.}

On the other hand, there is yet another natural pairing between
$H^p$ and $H_p$ where the latter is the singular homology, as
\bea
\int_{\Sigma^{(p)}} \omega_{(p)}\ ,
\eea
so we arrive at
\bea
&&{\rm dim}\, \cH_{L^2}^p (\cM_3) ={\rm dim}\, \cH_{L^2}^{6-p} (\cM_3) = {\rm dim}\, H_{6-p}(\cM_3)\ , \qquad p=0,1,2, \cr\cr
&&{\rm dim}\, \cH_{L^2}^3 (\cM_3)=0\ ,
\eea
for a toric Calabi-Yau 3-fold which is asymptotically conical. So,
the matter of BPS state counting for Abelian quivers with unit
rank at each node brings us to the homology counting for $p>3$.

Even before counting of $H_{p}(\cM_3)$, we already know
some universal facts. First, the toric Calabi-Yau 3-folds
in question are all algebraic manifolds, with only even homology
being non-trivial, meaning that only $H_{4}(\cM_3)$ and $H_{6}(\cM_3)$
are needed for extracting the entire $\cH_{L^2}^\star (\cM_3)$.
Second, the local Calabi-Yau has the top homology empty, $H_{6}(\cM_3)=0$
since one cannot possibly draw a top-dimensional simplex with
no boundary.  Therefore, one only needs to count $H_{4}(\cM_3)$,
whose dimension would also count  ${\rm dim} \cH_{L^2}^2 (\cM_3)
={\rm dim} \cH_{L^2}^{4} (\cM_3) $.

Perhaps, the simplest examples of local Calabi-Yau
3-folds are the conifold and the local $\IP^2$. These are, respectively, a
$\mathcal{O}(-1)\oplus \mathcal{O}(-1)$ bundle over $\IP^1$ and
a $\mathcal{O}(-3)$ line bundle over $\IP^2$. Given the homotopy
invariance of singular homology, we are immediately led to
\bea
H_\star ({\rm Conifold}) = H_\star (\IP^1)\ , \qquad H_\star ({\rm local}\; \IP^2) = H_\star (\IP^2) \,,
\eea
which says $\cH_{L^2}^\star ({\rm Conifold}) = 0$ according to (\ref{L_2thm}),
for example, and one finds
\bea
\cI_{\rm Conifold}=0\ , \qquad
\cI_{{\rm local}\; \IP^2} = -{\bf y}-{\bf y}^{-1}\ .
\eea
The conifold example actually belongs to the class we
already discussed in Section \ref{problem}, except that $K\ge N$
where the $L^2$ cohomology vanishes entirely given the same sign
of the FI parameter.

More relevant for us are the two well-known infinite classes of
toric Calabi-Yau 3-folds introduced in Section \ref{ex-quiver}.
The first set is the $Y^{p,q}$ family, whose toric diagram is
already given in Figure \ref{fig:ypq-toric-embed}.
For readers' convenience, we list its four external vertices here again,
\begin{align}\label{ypq-coord}
v_1 = (1,0), \quad v_2 = (0,0), \quad v_3 = (0,p), \quad v_4 = (-1, p-q)\,.
\end{align}
We take $q \neq p$, and also $q<p$ without loss of generality, such that
in our convention the toric diagram is convex.\footnote{A scaling choice
is also made in the geometry that treats the two $\IS^2$ in the base
of $Y^{p,q}$, and thus the two winding numbers $p$ and
$q$, differently in order to reach an $D=5$ $SU(p)_q$ theory.}

With this toric data, it is straightforward to compute their homology,
\begin{align}
\begin{split} \label{ypq-hdR}
{\rm dim}\, H_0 & = 1 \,, \\
{\rm dim} \,H_2 & =  p \,, \\
{\rm dim} \,H_4 = & \, \, p-1 \,, \\
{\rm dim} \,H_6 & = 0 \,.
\end{split}
\end{align}
In other words,
\begin{align} \label{ypq-hL2}
{\rm dim} \,\cH_{L^2}^2 ={\rm dim}\,\cH_{L^2}^4=p-1 \ ; \qquad  {\rm dim} \,\cH_{L^2}^p =0\ , \;\; p\neq 2,4 \,,
\end{align}
or
\bea
\cI_{Y^{p,q}}= (p-1)\times(-{\bf y}-{\bf y}^{-1}) \ .
\eea
Recall that M-theory compactified on a $Y^{p,q}$ 3-fold results
in $D=5$ $SU(p)_q$ gauge theory with the rank $p-1$.

Another infinite family of interest is the aforementioned $X^{p,q}$ family,
whose toric diagram is shown in Figure \ref{fig:xpq-toric-embed}. In addition to four vertices of $Y^{p,q}$ family, $X^{p,q}$ toric diagram has an extra vertex $v_5$ as noted in \eqref{xpq-toric-coord}.
We find the homology from the toric data as
\begin{align}
\begin{split} \label{xpq-h}
{\rm dim}\, H_0 & = 1 \,, \\
{\rm dim}\, H_2 = & \, \, p+1 \,, \\
{\rm dim}\, H_4 = & \, \, p-1 \,, \\
{\rm dim}\, H_6 & = 0 \, ,
\end{split}
\end{align}
again with vanishing odd homology. Their $L^2$ cohomology $\cH^\star_{L^2}$ is thus
the same as that of $Y^{p,q}$ Calabi-Yau families, since only $H_4$ matters,
such that
\bea
\cI_{X^{p,q}}= (p-1)\times(-{\bf y}-{\bf y}^{-1}) \ .
\eea
M-theory compactified on an $X^{p,q}$ 3-fold, with $p>q$ and appropriately
scaled, gives rise to a $D=5$ $SU(p)$ gauge theory with one
fundamental matter, also of rank $p-1$.

\subsection{What Have We Counted?} \label{interprete1D0}

Let us take a step back and understand the physical states
thus constructed from $ \cH_{L^2}^2 (\cM_3)\simeq \cH_{L^2}^{4} (\cM_3)$.
The quivers with unit rank at each node correspond to states
with a single KK charge only. What physical states in
these gauge theories are counted by this  counting?
The answer is obvious; in the Coulomb phase the only BPS particles
with neither the gauge charges nor the flavor charges are
the rank-many vector multiplets that belong to the Cartan
part of the gauge sector. Thus, the above song-and-dance ends
up counting these Cartan vector multiplets, each with unit KK charge
along the compactification circle $\IS^1$.

More concretely, given a quiver with the Higgs  moduli space $\cM_d$,
the K\"ahler 2-form $J$ defines the Lefschetz spin on differential
$n$-forms, via $SU(2)_L$ actions,
\bea
L_+[\omega_{(n)}]= J\wedge \omega_{(n)}\ ,\qquad L_-[ \omega_{(n)} ]= J \lrcorner \,\omega_{(n)}\ \ , \qquad L_3[\omega_{(n)}]= \frac{(n-d)}{2} \omega_{(n)} \ .
\eea
This $SU(2)_L$ has been identified as spatial angular momentum along
$\IR^3$, so we have counted the number of bound state pairs forming
spatial spin doublets. This spin content comes about from the relative
part of the dynamics, so it needs to be tensor-producted against
the standard half-hypermultiplet content from $U(1)_*$.
In other words, we would find vector multiplets out of
$\cH_{L^2}^\star (\cM_3)$, whose number equals ${\rm dim}\,H_{4}(\cM_3)$.

Indeed, the answers we found in Sec. \ref{secTHM} are all
such that one finds spin 1/2 multiplets, the number of which
always equals the purported rank of the corresponding $D=5$
supersymmetric field theory:
\bea
\cI_{(1,1,\dots,1)}({\bf y}) = {\rm rank}\times\left(-{\bf y}-{\bf y}^{-1}\right) \,.
\eea
So far, we have left out the center of mass motion along the
spacetime $\IR^{3+1}$, or equivalently the decoupled overall $U(1)$
vector multiplet, which supplies a half-hypermultiplet. The above
spin doublets  from the internal part combine with this half-hypermultiplet
from $\IR^{3+1}$ and elevate these states to rank-many vector multiplets.
In the end, we have recovered the Cartan vector multiplets with
a unit KK charge precisely via the proposed BPS quivers.

In some sense, through the above elaborate procedure, we did not
really dig up any new information. After all, the rank of
the gauge group should be equal to ${\rm dim}\cH_{L^2}^2(\cM_3)$,
which, either by the Poincar\'e duality, or equivalently by the
Lorentz symmetry, has to be the same as ${\rm dim}\cH_{L^2}^4(\cM_3)$.
What must be noted is, rather, how this recovery for unit KK modes
is achieved, only after a careful distinction between various
types of cohomology,
from the standard singular homology counting, via the theorem (\ref{L_2thm}):
We have counted ${\rm dim}\cH_{L^2}^2(\cM_3)$ by starting with
the usual singular homology counting of toric variety whose $H_4$
maps to $H^4(\cM_3)$ via the natural pairing and in turn to
$\cH_{L^2}^4(\cM_3)$ by the theorem of Section~\ref{secTHM},
bringing us to
\bea
{\rm dim}\cH_{L^2}^2(\cM_3)={\rm dim}\cH_{L^2}^4(\cM_3) = {\rm dim} H_4(\cM_3)\ ,
\eea
with all other $\cH_{L^2}^\star(\cM_3)$ empty, for local Calabi-Yau 3-folds.

\section{Towers of Pure D0 Branes}\label{nD0}

Let us now turn to the question of how one constructs the
entire KK towers for the Cartan vector multiplets; these
pure KK states should correspond to assigning a
common rank, say $n$, to all nodes of the BPS quiver.
For each KK momentum $n$, one expects to find precisely
rank-many spin doublets of supersymmetric ground states,
to be tensored by a universal half-hypermultiplet from
the decoupled $U(1)$. Recall that, despite various
difficulties and subtleties, the problem for $n=1$
at least remains that of the cohomology of the Higgs moduli
space. This is no longer true for $n>1$, as we see below.

\subsection{Cohomology of the Symmetric Orbifold is Irrelevant} \label{SnM}

Since $n=1$ quiver was a single D0 probe theory over the
Calabi-Yau $\cM_3$, the Higgs branch for $n=1$ is precisely $\cM_3$.
This motivates one to speculate that $n>1$ quiver may flow
down to a sigma model onto the $n$-th symmetric product of $\cM_3$, say,
\bea
\mathcal S^n\cM_3 = (\cM_3)^n/S_n \,.
\eea
Recall that the chiral multiplets in these BPS quivers are
either adjoint or bifundamental: At typical point in the Higgs
moduli space, the bifundamentals are turned on and reduce
the gauge groups to a single $U(n)$, namely, the common $U(n)_*$
that rotates all nodes simultaneously. Among this $U(n)_*$,
the overall $U(1)_*$ decouples from the rest of the dynamics
since no bifundamentals would be rotated by it, and its scalar superpartners serve
as the center of mass degrees of freedom along the spatial $\IR^{3}$.
This leaves behind the interacting $SU(n)_*$ theory, under
which all chirals transform as adjoint. This leads us to the
above symmetric orbifold.

One may hope that the (co-)homology of this symmetric orbifold is
the quantity to study. However, this is not correct.
There are two interrelated problems with this naive thought.
The first is that the theory actually flows to a different
symmetric orbifold,
\bea \label{nNLSM}
\cN_n\equiv \left[(\cM_3)^n\times (\IR^3)^n\right]/S_n=\IR^3\times
\left[(\cM_3)^n\times (\IR^3)^{n-1}\right]/S_n \,,
\eea
where $(\IR^3)^n$ comes from the Coulombic side. This is because,
at generic Higgs vev of bifundamental chirals and adjoint
chirals of the probe theory, the Cartan part of $SU(n)_*$
remains massless as well, so these Coulombic moduli can be turned
on simultaneously along with the chirals. Since we are
dealing with quiver GLSM with four supercharges, the former
translates to $3(n-1)$ free Coulombic moduli, i.e. relative
positions of $n$ D0 branes in the spatial $\IR^3$. Now, the orbifolding
group $S_n$ is the Weyl group of $SU(n)_*$, so it would act
on the Coulombic moduli and Higgs moduli simultaneously,
giving us the second factor in the rightmost  expression of (\ref{nNLSM}).

The second problem concerns, again, the obligatory $L^2$
condition along all moduli direction, except for
a single $\IR^3$ factor which is the center of mass. In other
words, the wavefunction must be $L^2$ on $\cN_n/\IR^3$.
There is a well-established generating function that
keeps track of Poincar\'e polynomial of a symmetric
orbifold, $\mathcal S^n\cM$, given that of $\cM$, due to
Macdonald \cite{MacDonald:1962}. However, as illustrated
in the Appendix \ref{sym}, one can understand the counting from this
generating function merely as that of $n$ identical particle
quantum mechanics with Bose statistics imposed.\footnote{
Please note that we are dealing with  $d=1$ quantum mechanics,
rather than $d=2$ theory which would have detected additional
localized states via twisted sectors.}

This means that any such $n$-particle states would be
plane-wave-like along each and every factor of $\IR^3$'s in
$\cN_n$. Even though one started with non-trivial $L^2$
wavefunctions found in $\cM_3$, there is no mechanism
at the level of the orbifold dynamics that forces two
or more of them to be confined along the Coulombic
part of the relative motion, $(\IR^3)^{n-1}$, for $n>1$.
As such, even if there are normalizable BPS states at
$n=1$, no other normalizable states can be found
for $n>1$ at the level of $L^2$ cohomology of $\cN_n/\IR^3$,
regardless of details of $\cM_3$.

\subsection{Entire KK Towers of the Cartan}\label{KKT}

If the $L^2$ cohomology of the low energy orbifold fails
to capture the desired BPS spectrum, we
must step back to the quiver GLSM. One might think that
topological invariants must be the same between these two
theories since one is merely the low energy limit of
the other, but in reality we just saw that this is too naive,
which is of course ultimately related to how the naive robustness
often fails in the presence of the continuum sectors \cite{Witten:1982df,Lee:2016dbm}.
As we saw above, the problem stems not only from the
noncompact Calabi-Yau but also from the Coulombic continuum.

A relatively simple example where GLSM and its low energy limit
thereof would offer different indices and twisted partition functions
is that multi D0 dynamics, namely maximally supersymmetric
$SU(n)$ Yang-Mills quantum mechanics.
The low energy limit is the symmetric orbifold $\IR^{9(n-1)}/S_n$
as the target, and the twisted partition function has been
computed in the past \cite{Lee:2016dbm}, with the unrefined limit being
\cite{Yi:1997eg,Sethi:1997pa,Green:1997tn,Moore:1998et}
\bea
\Omega_{\IR^{9(n-1)}/S_n} &=&  \frac{1}{n^2} \,, \cr\cr
\Omega_{SU(n)}&=& \sum_{p\vert n} \frac{1}{p^2} = 1+ \cdots+\frac{1}{n^2}\ ,
\eea
where, in the latter, $p$'s are divisor of $n$ including $1$.
Note how the two offer different twisted partition functions.
This goes against the naive thought that such quantities are
topological and thus should agree between UV and IR pictures.

We will see presently how the integral indices $\cI$, not just twisted
partition functions $\Omega$, differ between the two descriptions.
The twisted partition functions are rational, again due to
the asymptotically flat and gapless directions. The integral index
that counts the $L^2$ bound state requires a subtraction of
the latter continuum contribution, usually denoted as
$\delta\cI$, such that
\bea
\cI_{\IR^{9(n-1)}/S_n}&=&  \Omega_{\IR^{9(n-1)}/S_n} + \delta\cI_{\IR^{9(n-1)}/S_n} \,, \cr\cr
\cI_{SU(n)}  &=&\Omega_{SU(n)} + \delta\cI_{SU(n)} \,.
\eea
As one can show using the heat kernel method, the defect
terms $\delta\cI$'s are determined entirely by the asymptotic
dynamics of the either system \cite{Yi:1997eg,Sethi:1997pa}.

On the other hand, as one approaches the asymptotic region,
the orbifold becomes an ever more accurate approximation of the GLSM, since
all masses of off-diagonal parts, dropped in favor of the orbifold,
scale linearly with the Coulombic vev. This implies, given the
general nature of $\delta\cI$ as a boundary contribution
\cite{Yi:1997eg},
\bea
 \delta\cI_{SU(n)} & = & \delta\cI_{\IR^{9(n-1)}/S_n} +\cdots \ ,
\eea
where the ellipsis denotes terms that arise from hybrid sectors
with multiple partial bound states exploring the asymptotic
regions individually. This cascade of partial bound states
contributing to $\delta\cI$ was observed early on \cite{Green:1997tn,Kac:1999av}.

The $SU(n)$ theory is a little special in that the only contributing
continuum sectors are such that $n$ is divided equally to
$p$ identical $n/p$-particle bound states formed by $SU(n/p)$
interactions, whose mutual {\it asymptotic}
dynamics is governed by $\IR^{9(p-1)}/S_p$.
This implies
\bea\label{SUn}
 \delta\cI_{SU(n)} & = &\sum_{p\vert n}^{p\neq 1} \delta\cI_{\IR^{9(p-1)}/S_p} \times \cI_{SU(n/p)} \ ,
\eea
with which the only self-consistent answer in the end is
\bea
\delta\cI_{\IR^{9(n-1)}/S_n} &=&- \Omega_{\IR^{9(n-1)}/S_n} \,, \cr\cr
\cI_{\IR^{9(n-1)}/S_n}&=&0 \,, \cr\cr
\cI_{SU(n)}  &=&1 \,.
\eea
The last supports the existence of the M-theory as was
originally envisioned by Witten \cite{Witten:1995ex}.

One can understand this disparity between the GLSM
and the low energy orbifold limit more physically as follows:
As Polchinski demonstrated handsomely \cite{Polchinski:1999br},
a simple virial theorem, combined with supersymmetry, shows
that, on the supersymmetric ground state in question,
the expectation values of squared matrix elements are of
similar size between the ``diagonal" and ``off-diagonal" part
of $SU(n)$. Therefore, if one scales down the energy scale
toward infrared to render the off-diagonal components to
become very massive and ``ignorable," the support of the
wavefunction is shrunken further and further near the
origin along the Cartan directions as well. In the strict
geometric limit, such wavefunctions would then have
a vanishing support and  becomes undetectable by $d=1$ orbifold.

Interestingly, this prototype $SU(n)$ example is in fact all
one needs to address the infinite KK towers of the Cartan part
of the gauge sector that we are trying to establish. Why is so?
Although the orbifold $\cN_n$ is on its own irrelevant
for BPS state counting as we already saw, this moduli space is still
useful for visualizing where the desired bound states are supported.
Consider $(n,n,n,\cdots,n)$ quiver with the moduli space
\bea
\cN_n=\IR^3\times \left[(\cM_3)^n\times (\IR^3)^{n-1}\right]/S_n \,.
\eea
We have seen in the previous section that $n=1$ case produces
rank-many spin doublets along $\cM_3$, each of which is combined with a
half-hyper from the center of mass $\IR^3$ to form a vector
multiplet. Let us denote the wavefunction responsible for
these states,
\bea
\Psi_{\rm geometric}\
\eea
collectively. These states can be characterized as plane-wave-like
with half-hyper spin content along $\IR^3$ and as an $L^2$ harmonic
2-form/4-form pair along $\cM_3$.

Recall that, at generic point of the Coulomb phase of geometrically
engineered $D=5$ gauge theory, $\cM_3$ is a smooth manifold and
point-wise can be approximated by $\IC^3=\IR^6$. If we scale the
quiver theory such that the symmetric orbifold is better
and better approximation, the desired states must be that
unit-charged KK particles are clustered ever closer among
themselves, such that the above symmetric orbifold is better
and better approximated by
\bea\label{relative}
\cN_n\simeq [ \IR^3\times\cM_3]\times \left[(\IR^6)^{n-1}\times (\IR^3)^{n-1}\right]/S_n \,.
\eea
The former square bracket represents the center of mass
part of the dynamics while the latter is approximately
valid when the distances between the individual
bound states of the $(1,1,1,\cdots,1)$ sub-quivers
are relatively small compared to the curvature scale of $\cM_3$.

The relative part,
\bea
\left[(\IR^6)^{n-1}\times (\IR^3)^{n-1}\right]/S_n = (\IR^9)^{n-1}/S_n
\eea
can be regarded as the low energy limit of the $SU(n)_*$ subsector
we have introduced in Section~\ref{Q} for the orbifold examples.
However, we must emphasize again that one cannot expect to find
the relevant $L^2$ states at this geometric level. Rather, the
states in question, if any, would emerge only if we go back to the
full GLSM. Therefore, the desired wavefunctions may be
approximately factorized as
\bea\label{decomposition}
\Psi \simeq \Psi_{\rm geometric}\otimes \Psi_{SU(n)_*}
\eea
at least in the limit where the moduli space itself factorizes
approximately as in  (\ref{relative}).

Let us see more precisely how this happens. The two factors,
$\Psi_{\rm geometric}$ and the relative part of the wavefunction,
which is to be eventually replaced by $\Psi_{SU(n)_*}$, scale somewhat
differently under change of parameters of the quiver theories.
$\Psi_{\rm geometric}$ is represented by a $L^2$ harmonic form on
$\cM_3$, so its support is sensitive  to the scale of the
Calabi-Yau 3-fold in question, or equivalently to FI
constants of the BPS quiver. On the other hand, the relative
part of the wavefunctions is unaware of the FI constants
and scales with the gauge couplings. As one tunes the coupling
so that the symmetric orbifold limit is ever more accurate,
the support of the relative part of the wavefunction must become more
and more localized at the origin of the orbifold. On the other
hand, near this origin, the moduli space is increasingly similar
to (\ref{relative}), and the relative part of the quiver dynamics
approaches the $SU(n)_*$ theory.

One parameter that controls this process is the gauge
coupling, since the massive off-diagonal component, to be
dropped in the orbifold limit, would have the mass proportional
to the coupling.
This means that, by tuning the electric couplings of the
BPS quiver continuously (but never actually taking the limit),
one can make the support of the relative part much smaller than
the support of $\Psi_{\rm geometric}$.  Therefore there exists
a corner of the parameter space of the BPS quiver, where we can reliably
replace the BPS state counting problem by that of  (\ref{decomposition}).
This approximation is possible unless we are sitting
near a singular point of the center of mass $\cM_3$.
We are considering  generic point on the Coulombic moduli space of
$D=5$ theory, so $\cM_3$ would be smooth everywhere.
As such, (\ref{decomposition}) is reliable at least  for
the purpose of the index counting for BPS states,

We have uncovered, earlier in this section, the content of
$\Psi_{\rm geometric}$ as rank-many vector multiplets, it
remains to count $\Psi_{SU(n)_*}$ by going back to the
approximate $SU(n)_*$ theory. Fortunately, this more
difficult task is already performed since $SU(n)_*$ theory
is nothing but the maximally supersymmetric Yang-Mills theory
which we used above as an illustration. Our review above immediately
translates to
\bea\label{relativeindex}
\cI_{SU(n)_*}  &=&1\ ,
\eea
implying a unique supersymmetric $\Psi_{SU(n)_*}$ for each
integer $n$.  This should be contrasted against
\bea
\cI_{(\IR^9)^{n-1}/S_n}  &=&0\ ,
\eea
which supports our earlier claim that cohomology of
the symmetric orbifold does not have new localized
states ``created" due to the orbifolding. The gauge
theory ``resolution" of the orbifold singularity is
essential for $d=1$.

As such, the BPS state content of any orbifold
quiver is such that there are again precisely rank-many
vector multiplets at each and every KK charge $n$;
There is always a unique $\Psi_{SU(n)_*}$ state, so
the counting of the states are the same for all $n\ge 1$.
This translates to the following refined $L^2$ index
for $(n,n,\dots,n)$ quivers,
\bea
\cI_{(n,n,\dots,n)}({\bf y}) = {\rm rank}\times\left(-{\bf y}-{\bf y}^{-1}\right)
\eea
universally  where ``rank" refers to $D=4,5$ field theory in question.
The result is manifestly independent of the potential wall-crossing chambers,
since the relevant bound state comes from an effective $SU(n)_*$ theory
that has no FI parameter.

We have obtained this $SU(n)_*$ theory rather explicitly for
orbifold quiver theories; recall $U(n)_*$ and $SU(n)_*$ theories
we discussed  in Section~\ref{Q}. The two are related by how we factor
out the decoupled overall $U(1)$ and keep only the traceless
parts  of chirals $x,y,z$, which we called $x_*,y_*,z_*$. The cubic
superpotential,
\bea
W_* = {\rm tr}( x_*[y_*, z_*])
\eea
governs the low energy dynamics, rendering this $SU(n)_*$ theory
to become the maximally supersymmetric $SU(n)$ theory, leading
us to (\ref{relativeindex}), and the resulting rank-many vector
multiplets for each $n> 1$.

Furthermore, the spirit behind the
decompositions (\ref{relative}) and (\ref{decomposition}) clearly
holds for any smooth Calabi-Yau 3-fold $\cM_3$, as long as the
curvature length scale of the latter is taken sufficiently large
and the support of the relative part of the wavefunction is
controlled to be small by tuning the electric coupling.
As such, this result of KK towers of rank-many vector multiplets
still stands universally, although our line of thoughts cannot
be applied immediately to central, strongly interacting
regions of the Coulombic  moduli space of $D=5$ theory
in question.

Although we did this for positive $n$, the same for negative $n$ also
follows since nothing much changes upon $n\rightarrow -n$
except which half supersymmetries of the $D=5 $ theory are
preserved by the BPS state.

\section{Coulombic Counting} \label{Coulombic}

In recent years, computations of (refined) index for supersymmetric
gauged quantum mechanics have received renewed interest via the
localization method. While the most sweeping formulation of such
kind was given several years ago \cite{Hori:2014tda}, via Jeffrey-Kirwan residues,
this does not quite work for the BPS quivers here since the former
must assume that the superpotential is generic.

Although some global symmetries can be incorporated, allowing the
superpotential constrained further, non-Abelian global symmetries
are often not entirely reflected; Asymptotic isometries of local
Calabi-Yau's in question, cannot be fully incorporated into
such a computation. One might still hope that the Cartan part of such
isometries, whose chemical potentials do enter the computation,
suffices but one can see in various local Calabi-Yau examples, such as the
conifold quiver, the routine offered by these localization computation
often produces nonsensical results.

One alternative is an older routine of the Coulomb branch counting
\cite{Manschot:2010qz, Kim:2011sc, Manschot:2013sya, Manschot:2014fua},
whose mechanism by itself is insensitive to the superpotential of
the quiver since it keeps track of how BPS states are constructed
along the Coulombic moduli space; the superpotential data enters
obliquely, if crucially, via the single-center degeneracies treated
as input data, also known as the quiver invariant.

In this approach, the issues due to noncompact targets are split into
two different types. The subtleties due to the noncompact Higgs branch
are now all hidden in the quiver invariant \cite{Lee:2012sc,Lee:2012naa}
to be surmised through other routes. For our quivers,
in particular, the ubiquitous spectra of the neutral KK tower we
have already obtained restrict many of quiver invariants to vanish.
The other type due to the Coulombic continuum has a known
resolution in existing wall-crossing literature via a form of
multi-cover formulae \cite{Kontsevich:2008fj,Manschot:2010qz,Kim:2011sc,Lee:2016dbm}.
Rational invariants from the latter, again to be denoted as $\Omega$
here, have an explicit inversion formula and give $\cI$ as needed.
We will see how these work, first by revisiting multi-D0 states.

\subsection{Pure D0 Towers Revisited, with Quiver Invariants}

For general quiver theories, this multi-center approach is known to be
incomplete due to the so-called quiver invariants, which count
degeneracies of single-center BPS objects and are immune to wall-crossing
\cite{Bena:2012hf,Lee:2012naa,Lee:2012sc,Denef:2007vg}.
States counted by the quiver invariants  are localized at
the center of the Coulombic moduli space and instead spread along
the Higgs branch. This means that the missing superpotential data
which affects the Higgs branch only, would manifest in this Coulombic
approach via the quiver invariants, often denoted as $\Omega_S$ or
$\Omega_{\rm Inv}$ and generally rational if the quiver is not
primitive. We use $\cI_S$ for their integral and enumerative
counterpart for the sake of clarity.

One can view $\cI_S$ as an analog of the internal degeneracy, 2,
of an electron which is needed to construct Schr\"{o}dinger atoms.
A general solution \cite{Manschot:2010qz,Kim:2011sc,Manschot:2013sya}
to the wall-crossing formulae of Kontsevich and Soibelman
\cite{Kontsevich:2008fj} exists \cite{Sen:2011aa}, with the quiver
invariants as input data. On the other hand, $\cI_S\neq 0$ appears
generically in the black hole regime, and the known BPS spectra
of $D=4$ field theories tend to be consistent with  $\cI_S=0$
except for the $U(1)$ single-node quivers. Even with generic
superpotentials, as would be relevant for BPS black holes,
the quiver invariant tends to be absent when  all intersection
numbers are 2 or smaller \cite{Bena:2012hf,Lee:2012sc,Lee:2012naa}.

States counted by the quiver invariant are in particular dictated by
the F-term  embedding into D-term ambient \cite{Lee:2012naa}.
As such, the simpler, fine-tuned
superpotential tends to suppress these possibilities further.
A simple and non-trivial example is the triangle cyclic quiver
\cite{Denef:2007vg} with the common intersection number 3. For this,
 the Coulomb counting gives $\cI=-{\bf y}-{\bf y}^{-1}+\cI_S$
with $\cI_S$ unknown. With generic superpotentials, the quiver
theory flows to an elliptic curve such that $\cI=0$ and thus
$\cI_S={\bf y}+{\bf y}^{-1}$ \cite{Lee:2012sc}.
The same quiver with a fine-tuned superpotential with $SU(3)$
isometry,  on the other hand,  would flow to a local $\IP^2$,
for which we must have $\cI=-{\bf y}-{\bf y}^{-1}$ and $\cI_S=0$.

With the simplifying assumptions that $\cI_S=0$ except for the
elementary nodes, we have computed the Coulombic  index of several
$D=5$ BPS quivers with $(n,n,\dots,n)$ rank vectors, and successfully
reproduced,
\bea
\cI( n \hbox{ D0's};{\bf y})= {\rm rank}\times\left(-{\bf y}-{\bf y}^{-1}\right) \ ,
\eea
independent of $n$. Again, ``rank" refers to $D=4,5$ field theory
in question. Note that the mutual consistency between our counting
in Section \ref{D0} and \ref{nD0} and the current Coulombic counting requires
\bea
\cI_S( n \hbox{ D0's};{\bf y})= 0\ ,
\eea
in particular, and $\cI_S= 0$ for all nontrivial subquivers
as well.

The index $\cI(n\, \hbox{D0's};{\bf y})$ is independent of the
wall-crossing chambers. This chamber-independence has been
seen from the construction of Section \ref{D0} and \ref{nD0}; only
$\Psi_{\rm geometric}$ part can know about FI constants, $\zeta$,
but the degeneracy is insensitive to signs of $\zeta$ since the
topology remains independent thereof. The stability is also
numerically observed here, but more generally, the Coulombic
wall-crossing picture implies the same for all states whose
quiver rank vectors are in the kernel of the intersection matrix
\cite{Beaujard:2020sgs}.

When $n>1$ so that the quiver is no longer primitive, an important distinction
must be made between the true integral index $\cI$ and its rational cousin
$\Omega$. $\Omega$ could be directly computed by a Coulombic heat kernel
method \cite{Manschot:2010qz,Lee:2011ph}, from which $\cI$'s are extracted by inverting
\bea\label{eq:IandOmega}
\Omega(\Gamma;{\bf y})=\sum_{p\vert \Gamma}
\frac{{\bf y}-{\bf y}^{-1}}{p\cdot({\bf y}^p -{\bf y}^{-p})}\cdot \cI(\Gamma/p;{\bf y}^p) \ ,
\eea
where the sum is over divisors $p$ of the rank vector $\vec N$ of quiver $\Gamma$
and $\Gamma/p$ refers to a subquiver, whose rank is given by division of $\vec N$ by $p$. This sum
is exactly the kind of the rational structure, now refined, that we have
encountered for $\Omega_{SU(n)}=\cI_{SU(n)}-\delta \cI_{SU(n)}$ via
\eqref{SUn} for maximally supersymmetric $SU(n)$ theories.
The quiver invariant follows the same pattern, so $\Omega_S(  n \hbox{ D0's};{\bf y})=0$.

\subsection{Electrically Charged BPS States in the Weak Coupling Chamber}

For BPS states with charges coming from the gauge sector,
one should expect heavy wall-crossings to happen. Already
with such a relatively simple theory like $D=4$ $SU(2)$ $N_f=4$,
the wall-crossing pattern quickly becomes intractable,
so the situation with $D=5$ BPS quivers, with two extra nodes
relative to its $D=4$ counterpart, any sort of classification
of BPS states is not going to be practical. Any such attempt
of BPS state classification should be in practice accompanied
by well-motivated choices of the wall-crossing chamber.

One type of the chamber which might be viewed as a reference
is the weak coupling chamber. Note that the weak coupling limit
by itself does not preclude wall-crossings since dyons can
decay to a pair of other dyons easily when the rank of the group
is larger than one \cite{Bak:1999da,Gauntlett:2000ks,Stern:2000ie}.
For purely electric objects from the gauge group choice and
the matter content, however, there should not be any wall-crossing
in the weak coupling limit since  the theory would be defined
by such content. As such, the "weak coupling chamber" makes sense
for these elementary BPS objects with a relatively simple state
counting: Given how a KK tower would result from the $\IS^1$
compactification, we should expect to find
\bea
\cI(\Gamma+ n \hbox{ D0's}) = \cI(\Gamma)
\eea
for quiver $\Gamma$ whose net charge content corresponds to
an elementary state that defines the $D=5$ field theory in question.

Let us recall what we know about the weak coupling
regime of $D=4$ pure $SU(2)$ theory. The weak coupling
means very large $1/g^2$, so the monopole $M$ and the
dyon $D$, which together span all known BPS states of
$D=4$ pure $SU(2)$, would be much heavier than the
elementary charged vector boson $W$,
\bea
|Z_M| \gg |Z_W| \ , \qquad  |Z_D| \gg |Z_W|\ ,
\eea
while the central charges are related as $Z_W = Z_M+ Z_D$.
Drawing these three central charges in the upper half plane,
$Z_M$ and $Z_D$ have to be very large and point to almost
opposite of each other, while $Z_W$ should direct the halfway
between the two. On the other hand, FI constants measure
how far the central charges of each node, $Z_M$ and $Z_D$,
deviate from the total central charge, $Z_M+Z_D$ \cite{Denef:2002ru,Lee:2011ph}.
Therefore, the weak coupling chamber corresponds to a large
FI constant of some particular sign, opposite of the
other strong coupling chamber where $W$ is absent as a
state.

How does this generalize to $D=5$ pure $SU(2)$ theory? The question
comes down to understanding the central charges of the two additional
nodes. For $\mathbb{F}_0$ quiver
in Figure \ref{fig:f0-quiver-colored} of Appendix \ref{map}, the magnetic charge assignment to each node
is\footnote{This can be inferred from the brane charge of each node, which is summarized in Appendix \ref{map}.}
\bea
M\ , \quad -M\ , \quad - M\ , \quad M\ ,
\eea
so one might think that signs of the FI constants should
follow these signs. On a closer inspection, however, one
realizes that there are more charge components to the latter
two whose central charge contributions scale like $1/g^2$,
\bea
M\ , \quad -M\ , \quad I - M\ , \quad M - I\ ,
\eea
with $I$ denoting the instanton. In fact, $I-M$
corresponds to the so-called KK monopole, which is
present whenever a $D\ge 4$  gauge theory is put on
a circle, in such a way that the full instanton $I$
is recovered from $M+(I-M)$ \cite{Lee:1997vp}.

The last node with magnetic charge $M-I$ carries
the KK charge but the latter's central charge contribution $\sim 1/R$
scales differently, as it does not get renormalized
with the gauge coupling renormalization. Pure electric
central charge, which should be added to the various
nodes, would also scale with $1/R$, in the regime where
the bulk of their mass comes primarily from the $\IS^1$
holonomy vev. In such a weak coupling regime, neither
of these additional types of charges is the leading
contribution to the central charges of the elementary nodes.

\begin{figure}[ht]
\begin{center}
\resizebox{\hsize}{!}{
\includegraphics[height=6cm]{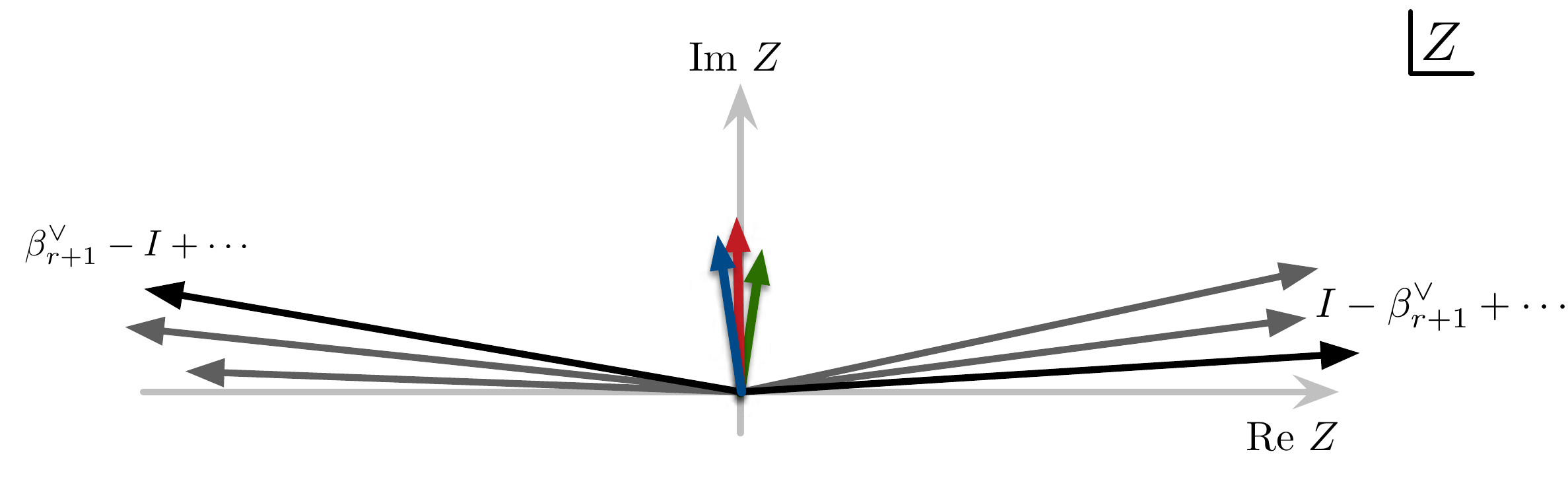}
}
\caption{
\small In the weak coupling limit, the elementary nodes are categorised
into three classes according to their central charges. The first, short
arrows in the middle, are weak
coupling objects, such as quarks and charged vectors. The second are
rank-many objects with positive magnetic charges, $\beta_{i \leq r}^\vee$,
plus the KK monopole, $I-\beta^\vee_{r+1}$. These are long arrows to the
right. The third group with the opposite magnetic/instanton charges
points toward the left. If the total central charge consists only of
electric, flavor, and KK charges, the second and the third classes
of nodes would be equipped with large positive and large negative
FI constants, respectively.
\label{fig:z-plane}}
 \end{center}
\end{figure}

Therefore, in the weak coupling limit achieved by a small
$\IS^1$ radius $R$ and large Wilson line expectation
values $\sim 1/R$, the node 1 and the node 3 each
should then have large positive FI constant while node 2
and node 4 should have large negative FI constant.
Since the FI constants measure deviation of the
node central charges away from the total central charge of the
quiver, this characterization is robust whenever the total
charge of the quiver has neither magnetic nor instanton components.

This story  generalizes to all higher rank gauge theories
with or without flavors. For example, consider a $D=5$
theory with a rank $r$ simple gauge group and $N_f$ fundamental
hypermultiplets; the number of nodes for the BPS quiver is $2r+2+N_f$.
Of these, $N_f$ captures the matter electric charges and bare
flavor masses thereof, while $2r+2$ nodes are analog of the
monopole node and the dyon node of pure $D=4$ $SU(2)$ BPS quiver.

With simple roots $\{ \beta_i\}$, $2r$ nodes of $D=4$ BPS
quiver would have magnetic charges
with the dual root $\beta_i^\vee$ \cite{Fraser:1996pw}
\bea
\beta^\vee_1\ , \quad -\beta^\vee_1\ , \quad \beta^\vee_2 \ , \quad -\beta^\vee_2\ , \quad \dots \quad, \quad \beta^\vee_r \ ,\quad - \beta^\vee _r \ .
\eea
The rest of the nodes, if flavor hypermultiplets
are present, corresponds to these elementary matter
BPS states \cite{Alim:2011kw}.

On top of these, $D=5$ BPS quiver requires two additional
nodes which, respectively, should carry the magnetic and
the instanton charges as \cite{Lee:1997vp,Lee:1998vu}
\bea
I- \beta^\vee_{r+1}\ , \qquad   \beta^\vee_{r+1} - I \ ,
\eea
where $\beta^\vee_{r+1} \equiv \sum a_i^\vee\beta^\vee_i$ is the
highest dual root. Recall that $h^\vee=\sum a_i^\vee$ is the
dual Coxeter number and
$\{\beta_1^\vee, \cdots, \beta_r^\vee,-\beta_{r+1}^\vee\}$
span the affine Dynkin diagram of the dual group.

Again, the weak coupling chamber translates to large
positive FI constants for nodes with magnetic charges,
$\beta_{i \leq r}^\vee$ and $- \beta^\vee_{r+1}$, and
large negative FI's for nodes with magnetic charges,
$-\beta_{i \leq r}^\vee$ and $\beta^\vee_{r+1}$. With such
an assignment, one should be able to recover, from the
$D=5$ BPS quiver, the expected KK towers of elementary
fields, such as charged vector mesons, collectively
denoted as W, and quarks, also collectively Q,
if fundamental flavors are added; that is\footnote{Similar
observations have been made in the past. See
Ref.~\cite{Bousseau:2019ift} for example where, in a large
volume limit of the local $\IP^2$, $n$-independence of
the Donaldson-Thomas invariants for $k$D2-$n$D0 were
conjectured, and demonstrated for $k\le 4$.}
\bea
&&\cI(\hbox{W} + n \hbox{ D0's}) = \cI(\hbox{W})=-{\bf y}-{\bf y}^{-1} \ ,\cr\cr
&&\cI(\hbox{Q}+ n \hbox{ D0's}) = \cI(\hbox{Q})=1 \ .
\eea
We have tested this for various BPS quivers we have discussed.
See Table \ref{tab:index}, where
we scanned up to 11 particle problems and also confirmed
that the index is robust under small variation of FI
constants.

\vskip 1cm

\begin{table}[ht!]
\begin{center}
\begin{adjustbox}{width=0.93 \textwidth,center=\textwidth}
\begin{tabular}{ |c|c||c|c|}
\hline
$\vec N$ & { BPS states} &  $\cI$ & FI constants  \\
 \hline
 \hline
 \multicolumn{4}{|c|}{$\mathbb{F}_0$, \, pure $SU(2)$, \, $\zeta_\text{w}= C (1,-1,1,-1) $} \\
 \hline
 $(1 , 1 , 0 , 0)$ & W  & $-{\bf y}-{\bf y}^{-1}$ & $\zeta_\text{w}$ \\
 $(2 , 2 , 1, 1)$ & W + KK &  $-{\bf y}-{\bf y}^{-1}$ & $ \zeta_\text{w} \pm \epsilon ( 1,-1, -1,1)  $ \\
 $(3 , 3 , 2, 2)$ & W + 2 KK & $-{\bf y}-{\bf y}^{-1}$ & $\zeta_\text{w} \pm \epsilon ( 1,-1, -1,1) $ \\
 \hline

 \multicolumn{4}{|c|}{$\mathbb{F}_1$, \, pure $SU(2)_{\pi}$, \, $\zeta_\text{w} =C (1,-1,1,-1)$} \\
 \hline
 $(1 , 1 , 0 , 0)$ & W & $-{\bf y}-{\bf y}^{-1}$ & $\zeta_\text{w}$ 
 \\
 $(2 , 2 , 1, 1)$ & W + KK  & $-{\bf y}-{\bf y}^{-1}$ & $ \zeta_\text{w} \pm \epsilon (1,-1,-1,1)  $ \\
 \hline

 \multicolumn{4}{|c|}{$dP_2$, \, $SU(2)$ with $N_f=1$, \, $\zeta_\text{w} = C (1, -1, 1, 0, -1) $} \\
 \hline
 $(1 , 1, 0 , 0 , 0)$ & W & $-{\bf y}-{\bf y}^{-1} $ & $\zeta_\text{w} $\\

 $(2 , 2, 1 , 1, 1)$ & W + KK & $-{\bf y}-{\bf y}^{-1}$ & $\zeta_\text{w} \pm \epsilon (1,-1,-1,1,0) $ \\

 $(1, 1, 1, 2, 1)$ & Q + KK & 1 & $\zeta_\text{w} \pm \epsilon (1,-1,-1,0,1) $ \\
 $(2, 2, 2, 3, 2)$ & Q + 2 KK & 1 & $\zeta_\text{w} \pm \epsilon (1,-3,2,0,0)  $ \\
 \hline

 \multicolumn{4}{|c|}{$dP_3$, \, $SU(2)$ with $N_f=2$, \, $\zeta_\text{w} = C (1,-1,0,0,1,-1) $} \\
 \hline
 $(1, 1, 0, 0, 0, 0)$ & W & $-{\bf y}-{\bf y}^{-1}$ & $\zeta_\text{w}$\\
 $(2, 2, 1, 1, 1, 1)$ & W + KK & $-{\bf y}-{\bf y}^{-1}$ & $ \zeta_\text{w} \pm \epsilon (1,0,0,0,2,-4) $\\
 $(1, 1, 1, 2, 1, 1)$ & $\text{Q}_2$ + KK & $1$ & $ \zeta_\text{w} \pm \epsilon (-2,-2,0,2,1,-1)  $\\
 \hline
 \multicolumn{4}{|c|}{$Y^{3,0}$, \, pure $SU(3)_0$, \, $\zeta_\text{w} = C (1,-1,1,-1,1,-1) $} \\
 \hline
 $(1, 1, 0, 0, 0, 0)$ & $\text{W}_1$ & $ -{\bf y}-{\bf y}^{-1} $ & $\zeta_\text{w}$ \\
 $(0, 0, 1, 1, 0, 0)$ & $\text{W}_2$ & $ -{\bf y}-{\bf y}^{-1} $ & $\zeta_\text{w}$ \\
 $(2, 2, 1, 1, 1, 1)$ & $\text{W}_1$ + KK & $-{\bf y}-{\bf y}^{-1}$ & $ \zeta_\text{w} \pm \epsilon (1,-2,1,-2,-1,4) $ \\
 $(1, 1, 2, 2, 1, 1)$ & $\text{W}_2$ + KK & $-{\bf y}-{\bf y}^{-1} $ & $\zeta_\text{w} \pm \epsilon (1,-1,-1,1,0,0) $ \\
 \hline

 \multicolumn{4}{|c|}{$Y^{3,1}$, \, pure $SU(3)_1$, \, $\zeta_\text{w} = C (1,-1,1,-1,1,-1) $} \\
 \hline
 $(1, 1, 0, 0, 0, 0)$ & $\text{W}_1$ & $-{\bf y}-{\bf y}^{-1} $ & $\zeta_\text{w}$ \\
 $(0, 0, 1, 1, 0, 0)$ & $\text{W}_2$ & $-{\bf y}-{\bf y}^{-1} $ & $\zeta_\text{w}$ \\
 $(2, 2, 1, 1, 1, 1)$ & $\text{W}_1$ + KK & $-{\bf y}-{\bf y}^{-1} $ & $ \zeta_\text{w} \pm \epsilon (1,-2,-1,-2,1,4) $ \\
 $(1, 1, 2, 2, 1, 1)$ & $\text{W}_2$ + KK & $-{\bf y}-{\bf y}^{-1} $ & $\zeta_\text{w} \pm \epsilon (1,-1,-1,1,0,0)$ \\
 \hline
\end{tabular}
\end{adjustbox}
\end{center}
\caption{\small Refined Witten index of BPS quivers with no net magnetic or
instanton charges, in the weak coupling chamber. We took $C\gg 1$ and $\epsilon\sim O(1)$.
Quiver invariants $\cI_S$ are all assumed to vanish except for elementary nodes.  \label{tab:index}}
\end{table}

\section{Summary}

We explored $D=5$ BPS quivers for field theories \cite{Closset:2019juk}
that are geometrically engineered, from M-theory, over a toric local
Calabi-Yau 3-fold and further compactified on a circle
$\IS^1$. The basic building blocks include those for the
corresponding $D=4$ BPS quiver, while instantons and KK modes
also enter via the additional nodes of the quiver. Altogether,
the BPS quiver can be read off from the D0 probe quiver,
which in turn can be deduced from older stories of D3
quivers \cite{Feng:2000mi}, in type IIB, that probe
the same Calabi-Yau 3-fold.

These BPS quivers are equipped with fine-tuned superpotentials
which are unavoidable for local Calabi-Yau's and render some of
the usual path-integral machinery for computing twisted partition
functions ineffective. In this note, we
delineated the computational issues, and addressed BPS counting
problems with emphasis on the KK towers.
Given the heavy wall-crossing patterns, cataloging of all
wall-crossing chambers is all but impossible, and we mostly
concentrated on two simplest classes of states: neutral vector
multiplets with KK charges, and electrically charged states
with KK charges in the weak coupling chamber.

The former, corresponding to the Cartan part of the gauge multiplets,
is the robust part of $D=5$ BPS spectra, entirely free of wall-crossing.
For states with a unit KK charge, the counting problem does
reduce to a geometric one, but with the $L^2$ condition entering
crucially. A universal routine for extracting $L^2$ cohomology, from
the more accessible ordinary de Rham cohomology, or equivalently
singular homology, was outlined \cite{Hausel:2002xg}, and we performed
the computation for several examples. One notable fact is that
the resulting refined indices are all symmetric Laurent polynomials,
as is necessary from the field theory viewpoint and contrary
to the naive homology counting.

For higher KK states of these neutral vector multiplets, governed
by non-Abelian quivers with rank vector $(n,n,\dots,n)$,
we show how the geometric cohomology approach fails entirely,
forcing us to consider the full gauged dynamics. Fortunately, multi-D0
wavefunctions in flat spacetime \cite{Yi:1997eg,Sethi:1997pa,Lee:2016dbm},
well-known from the famed M-theory/type IIA duality \cite{Witten:1995ex},
constitute the difficult relative part of $n>1$ BPS states, allowing
us to reconstruct the entire KK towers in the limit of small internal
curvatures.

BPS states with electric charges do suffer heavy wall-crossings,
and more so with nonzero KK charges, but admit a universal notion
of the weak coupling chamber. This should be contrasted against
magnetically charged BPS states which generically wall-cross
even in such a weak coupling regime \cite{Gauntlett:2000ks,Stern:2000ie}.
For these weak-coupling states, we relied on the Coulombic approach
\cite{Manschot:2010qz} where the intricacies due to the fine-tuned
superpotential are expected to enter via the so-called quiver invariants
\cite{Lee:2012sc,Lee:2012naa} that are input data for such multi-center
approach \cite{Manschot:2013sya}. Assuming quiver invariants are all
trivial except for the elementary node states in half-hypermultiplets,
we numerically recovered the anticipated KK towers. The same
method was also applied for the above neutral KK towers with equal
success.

We have recovered BPS states that are immediately expected from
the weak coupling content of the $D=5$ theories, and, as such,
this note could be viewed as a first-principle  confirmation of
the proposed $D=5$ BPS quivers, with heavy emphasis on the field
theory. Some of issues and difficulties  we pointed out for enumerative
(refined) indices are relevant whenever one considers theories
with noncompact targets, such as local Calabi-Yau's, so the
note can also be considered a cautionary lesson.

In particular, the Donaldson-Thomas (DT) invariants have been counted
for certain limited collections of D-branes or for simpler types
of Calabi-Yau's \cite{Behrend:2009dc,Morrison:2011bc,Morrison:2011rz}.
With noncompact Calabi-Yau's, one again finds the resulting motivic DT invariant
$\Omega_{\rm DT}({\bf y})$ not necessarily symmetric
under ${\bf y}\rightarrow {\bf y}^{-1}$.
On the other hand, recall how we reached at the $L^2$ index $\cI$
from such asymmetric twisted partition functions or from the vanilla
cohomology data, also asymmetric, in Sections~\ref{problem}~and~\ref{D0};
A natural, if na\"ive, question is whether taking the common part
of $\Omega_{\rm DT}({\bf y})$ and $\Omega_{\rm DT}({\bf y}^{-1})$
blindly might produce a sensible $L^2$ index.

The motivic DT invariant for $n$ D0's can be read off from
those for D6-$n$D0 in Ref.~\cite{Behrend:2009dc}, via wall-crossing,
and appears to have a universal form \cite{Manschot:2010qz} as\footnote{We thank Boris Pioline for
explaining these results.}
\bea
\Omega_{\rm DT}(n\hbox{ D0's};{\bf y}) = - P({\bf y})/{\bf y}^3
\eea
with the ordinary Poincar\'e polynomial $P$ of the Calabi-Yau in
question. Note the $n$-independence, just as with $\cI(n\hbox{ D0};{\bf y})$.
Given how $\cI$ for $n=1$ was extracted from the asymmetric
Poincar\'e polynomial of the local Calabi-Yau's in Section~\ref{D0},
we arrive at, in retrospect,
\bea
\cI(n\hbox{ D0's};{\bf y}) =
\Omega_{\rm DT}(n\hbox{ D0's};{\bf y})\cap \Omega_{\rm DT}(n\hbox{ D0's};{\bf y}^{-1})\ ,
\eea
where we used  $\cap$ as a shorthand for taking of the common
part of the two Laurent polynomials. It would be interesting
to see if this simple pattern between $L^2$ indices and the DT
invariants is applicable further.

A recent related work \cite{Beaujard:2020sgs} studied ground states
for $D=5$ BPS quivers with a numerical multi-center approach.\footnote{
Other recent studies on $D=5$ BPS spectra include 
Refs.~\cite{Banerjee:2018syt,Banerjee:2019apt,Bonelli:2020dcp}.} 
The emphasis there was on magnetically charged states in the so-called
canonical chamber, however, as appropriate for the Vafa-Witten
invariants that stem from the worldvolume theory of D4 branes wrapped
on 4-cycles. As such, the question of neutral and electric KK
towers was somewhat orthogonal to their interests, and left 
ambiguous.

One useful byproduct encountered along the way, both here and in
Ref.~\cite{Beaujard:2020sgs}, is how the Coulombic approach was
trustworthy. As long as $\cI_S=0$ continues to hold, the Coulombic
counting offers a routine that bypasses the complications due to
the fine-tuned superpotential and the accompanying $L^2$ condition.
How far this can be pushed remains to be seen, however.

\section*{Acknowledgement}

We are indebted to  the collaboration of J. Manschot, B. Pioline, and A. Sen,
who made a Coulombic multi-center code  publicly available. We have relied on
the routine for numerical computations in Section \ref{Coulombic}. We also thank
Sung-Soo Kim and Boris Pioline for useful discussions.  This work
is also supported by KIAS Individual Grants (PG076901 for ZD, PG071301
and PG071302 for DG, PG005704 for PY) at Korea Institute for Advanced Study.

\appendix

\section{Poincar\'e Polynomial and the Symmetric Product} \label{sym}

The rank-$n$ symmetric product of a manifold $\mathcal{M}$ is, by definition,
\begin{align}
    \mathcal{S}^{n} \mathcal{M} \equiv (\cM \times \cdots \times \cM) / S_n\,,
\end{align}
where we first take $n$-fold direct product of $\cM$ then mod out the symmetric group action via permuting $n$ elements.
When the manifold $\cM$ has complex dimension one, this space is known to be smooth.
On the other hand, if the dimension is greater than one the latter has singularities.
In the math literature, there exists a beautiful formula which expresses the cohomology of $\mathcal{S}^{n} \cM $ in terms of those of $\cM$. To state the result,
recall that the Poincar\'e polynomial $P_{\cM} (t)$ of $\cM$ is defined to be
\begin{align}
    P_{\mathcal{M}} (t) = \sum_{i \geq 0} b_i(\mathcal{M}) \cdot t^i\,,
\end{align}
with $b_i(\mathcal{M})$ the $i$-th Betti number of $\mathcal{M}$. Moreover, we encapsulate all the Poincar\'e polynomials of $\mathcal{S}^{n} \mathcal{M}$ into another generating function $\Pi_{\mathcal{M}}$,
\begin{equation}
    \Pi_{\cM}(t;y) \equiv  \sum_{n \geq 0} P_{\mathcal{S}^{n} \cM} (t) \cdot y^n\,.
\end{equation}
Then the formula takes the following form \cite{MacDonald:1962},
\begin{align}
\label{poincare-sympro}
    \Pi_{\cM} (t;y) =  \exp\left(\sum\limits_{r > 1}P_{\cM} (t^r) \, \frac{y^r}{r}\right) \,,
\end{align}
where we assigned $P_{\mathcal{S}^0 \cM} (t) = 1$ for convention.

The physical picture behind this formula is clear.
The generating function $\Pi_{\mathcal{M}}$ in \eqref{poincare-sympro} captures Bose-symmetrized cohomology out of full cohomology of product manifold $\mathcal{M}^{\otimes n}$,
where the latter is made by taking tensor products among cohomology elements of original manifold $\mathcal{M}$.
For instance, the Poincar\'e polynomial of the second and third symmetric product of $\cM$ is written in terms of $P_\cM$ as follows,
\begin{align}
\begin{split}
    P_{\mathcal{S}^2 \cM} (t) & =  \frac{P_\cM (t)^2}{2! } + \frac{P_\cM (t^2) }{2} \,, \\
    P_{\mathcal{S}^3 \cM } (t) = & \frac{P_\cM (t)^3}{3!} + \frac{P_\cM (t) P_\cM (t^2)}{2} + \frac{P_{\cM}(t^3)}{3} \,,
\end{split}
\end{align}
which manifests the construction of two- and three-particle partition function of non-interacting bosons out of its single-particle partition function.

This means that there are no fundamentally new states associated with the
symmetric orbifold. All states in the cohomology of $\mathcal{S}^{n} \mathcal{M} $
are constructed as tensor products of $n$ number of states that belong
to $H^\star(\cM)$, only to be symmetrized via the bosonic statistics. This is
of course natural since the $n$-particle Hamiltonian would be merely
a sum of $n$ mutually independent 1-particle Hamiltonian on
$\mathcal{S}^{n} \mathcal{M}$. It follows that the $L^2$ property of
these states classified by $H^\star(\cM^n/S_n)$ would follow immediately from
that of states classified by $H^\star(\cM)$.

The final piece of information for us is that
\bea
\cM=\IR^3 \times \cM_3\ ,
\eea
and that the $L^2$ property should be demanded over
\bea
\cM^n/\IR^3=\IR^{3(n-1)} \times (\cM_3)^{n} \ .
\eea
However, any non-trivial element of $H^\star(\cM)$, say for $n=1$,
would be uniformly supported along $\IR^3$ part of $\cM$. For $n>1$,
the bosonized wavefunction would immediately fail the $L^2$ condition,
due to this uniform spread along $\IR^3$ part of $\cM$. As such, no $L^2$
cohomology $\cH_{L^2}$ can exist, except for $n=1$.

\section{Internal Cycles and Fractional Branes} \label{map}

This appendix summarizes the brane charge assignment to nodes in each BPS quiver.
The following geometries are considered: local $\mathbb{P}^2 \,,
\mathbb{F}_0 \,, \mathbb{F}_1 \simeq dP_1 \,,$ $dP_2 \,, dP_3 \,,$ $Y^{3,0}$ and $Y^{3,1} \,. $
For basis choice of internal cycles, we refer readers to \cite{Closset:2019juk}.
According to their convention, we use $[\mathbf{E}_i]$ for four-cycles and
$[\cC_i]$ for two-cycles.

The corresponding BPS objects are monopole strings from M5 branes wrapped on the four-cycles, elementary charged fields from M2 branes wrapped
on the two-cycles, and D0 branes for $[\text{pt}]$.  In particular, the scaling of
the local Calabi-Yau is such that some of two cycles wrapped by M2 branes also
carry instanton charges. For all our examples with simple gauge groups, there
is only one such two-cycle, which we indicate in red.

We discuss local $\mathbb{P}^2$ theory first.
The D0 probe quiver of this theory is drawn in Figure \ref{fig:z3-quiver}.
Its fractional brane charges come as follows.
\begin{align} \label{p2-frac-brane}
    K[\cE_1] = [\mathbf{E}_0 ] \,, \quad K[\cE_2] = - 2 [\mathbf{E}_0] + [\cC_1] \,, \quad
    K[\cE_3] = [\mathbf{E}_0] - [\cC_1] + [\text{pt}] \,.
\end{align}

\begin{figure}[ht]
\begin{center}
\resizebox{0.5\hsize}{!}{
\includegraphics[height=6cm]{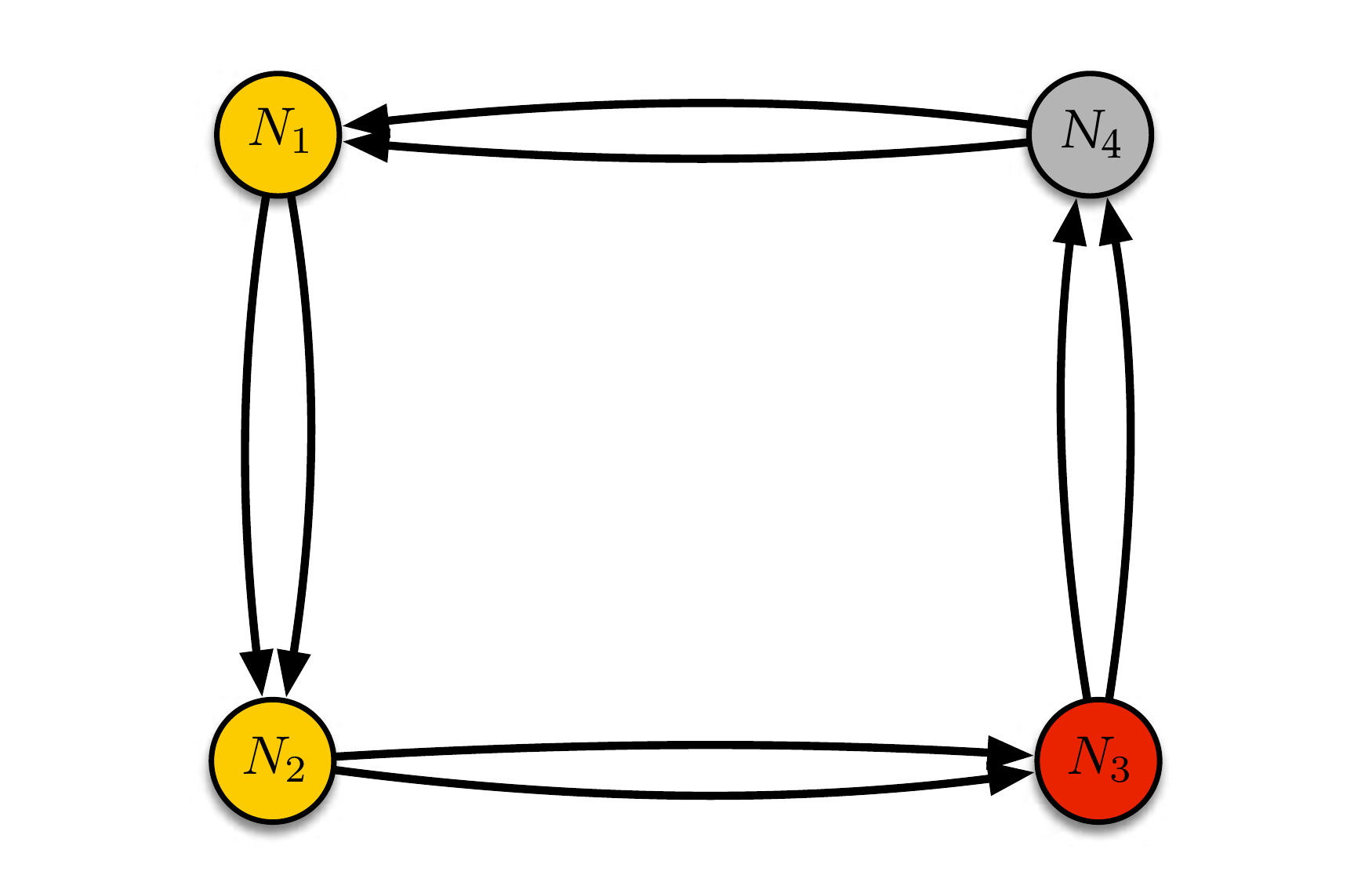}
}
\caption{\small
The BPS quiver of $\mathbb{F}_0$ theory. Yellow nodes and the arrows among them constitute the BPS quiver of $D=4$ theory.
The red node carries the KK monopole charge while the gray one carries unit KK charge as well as
negative of the KK monopole charge. \label{fig:f0-quiver-colored}}
 \end{center}
 \end{figure}

For $\mathbb{F}_0$ geometry, which engineers pure $SU(2)$ gauge theory in $D=5$,
the dictionary is as follows,
\begin{align} \label{f0-frac-brane}
\begin{split}
    K[\cE_1] = [\mathbf{E}_0 ] \,, \quad \qquad & \qquad \qquad \, K[\cE_2] = -  [\mathbf{E}_0] + [\cC_1] \,, \\
    K[\cE_3] = - [\mathbf{E}_0] - [\cC_1] & + [\textcolor{red}{\cC_2}] \,, \quad
    K[\cE_4] = [\mathbf{E}_0] - [\textcolor{red}{\cC_2}] + [\text{pt}] \,.
\end{split}
\end{align}
The probe quiver is given in Figure \ref{fig:f0-quiver-colored} for readers' convenience.

\begin{figure}[ht]
\begin{center}
\resizebox{0.5\hsize}{!}{
\includegraphics[height=6cm]{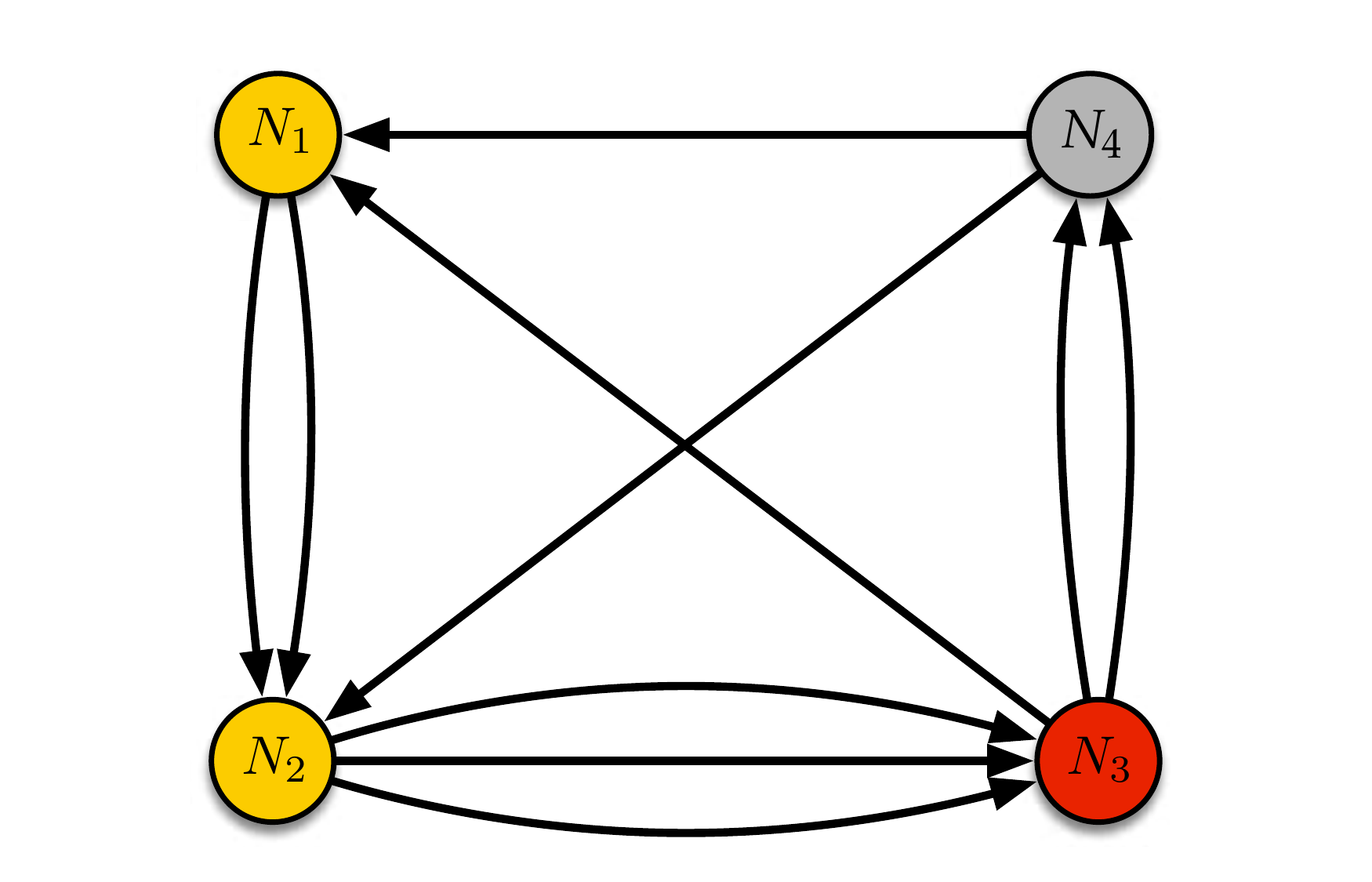}
}
\caption{\small
The BPS quiver of $\mathbb{F}_1$ theory.
\label{fig:f1-quiver}}
 \end{center}
 \end{figure}

$\mathbb{F}_1 \simeq dP_1 $ geometry, which engineers $SU(2)_{\pi}$ theory,
with a discrete theta angle associated to $\pi_4(SU(2))$ turned on,
has the following map,
\begin{align} \label{f1-frac-brane}
\begin{split}
    K[\cE_1] = [\mathbf{E}_0 ] \,, \quad \qquad & \qquad \qquad \, K[\cE_2] = -  [\mathbf{E}_0] + [\cC_1] \,, \\
    K[\cE_3] = - [\mathbf{E}_0] - [\cC_1] & + [\textcolor{red}{\cC_2}] \,, \quad
    K[\cE_4] = [\mathbf{E}_0] - [\textcolor{red}{\cC_2}] + [\text{pt}] \,.
\end{split}
\end{align}
Its quiver is drawn in Figure \ref{fig:f1-quiver}.

$dP_2$ geometry renders the next map.
This local CY engineers $D=5$ $SU(2)$ gauge theory with $N_f=1$ flavor. Its probe quiver is drawn in Figure \ref{fig:dp2-quiver}.
\begin{align} \label{dp2-frac-brane}
\begin{split}
    K[\cE_1] = [\mathbf{E}_0 ] \,, \quad K[\cE_2] & = -  [\mathbf{E}_0] + [\cC_3] + [\cC_4] \,, \quad
    K[\cE_3] = - [ \cC_3] \,, \\
    K[\cE_4] = & - [\mathbf{E}_0] - [\cC_3] - [\cC_4] + [\textcolor{red}{\cC_5}]  \,, \\
    K[\cE_5] = & \, [\mathbf{E}_0] + [\cC_3] - [\textcolor{red}{\cC_5}] + [\text{pt}] \,.
\end{split}
\end{align}

\begin{figure}[ht]
\begin{center}
\resizebox{0.5\hsize}{!}{
\includegraphics[height=6cm]{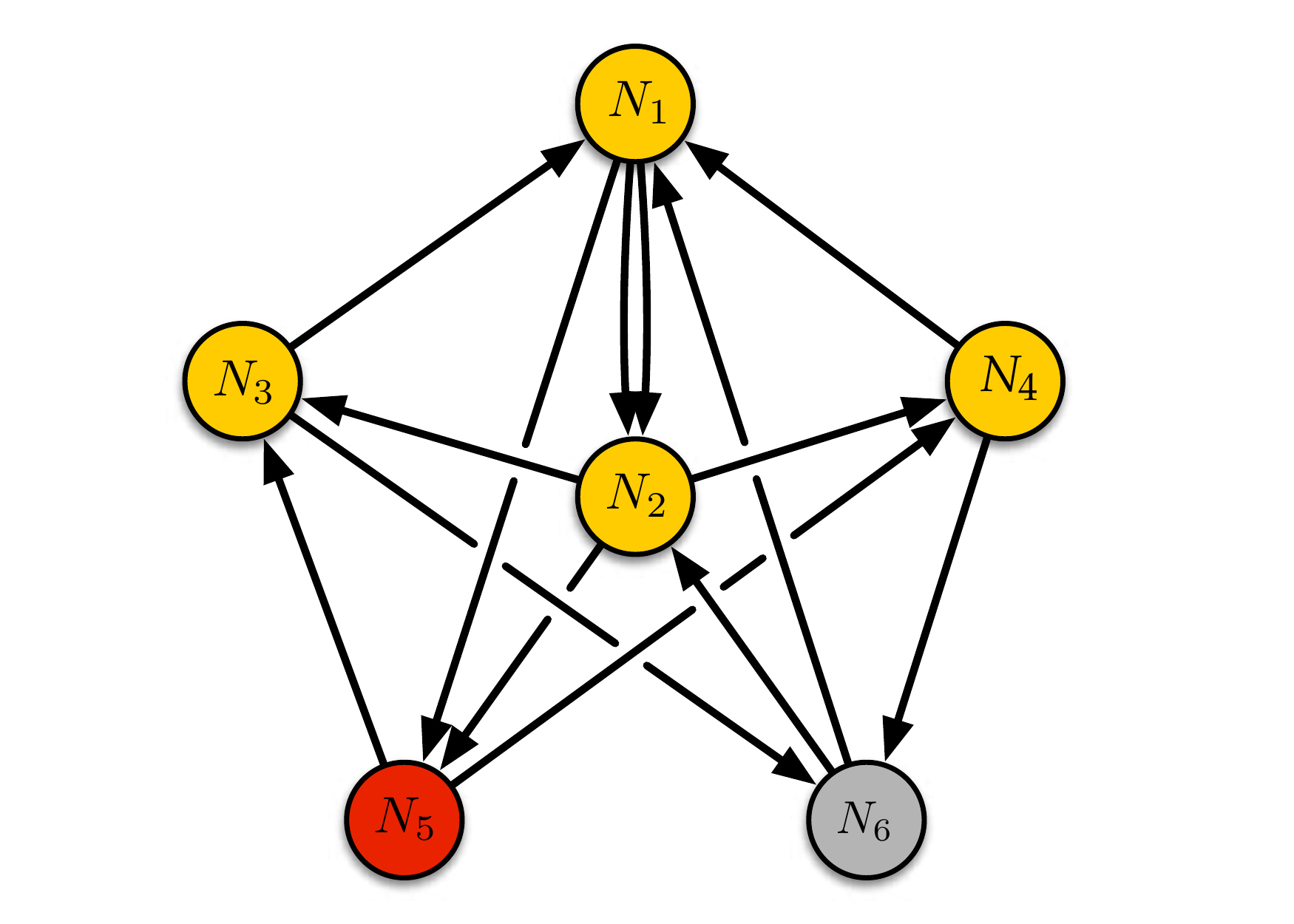}
}
\caption{\small
The quiver diagram of $dP_3$ theory.\label{fig:dp3-quiver}}
 \end{center}
 \end{figure}

$dP_3$ theory makes the fifth table. This local CY engineers $D=5$ $SU(2)$ gauge theory with $N_f=2$ flavor.
\begin{align} \label{dp3-frac-brane}
\begin{split}
    K[\cE_1] = [\mathbf{E}_0 ] \,, \, \, \, & \qquad K[\cE_2] = - [\mathbf{E}_0] + [\cC_3] + [\cC_4 ] \,, \\
    K[\cE_3] = -[\cC_1] \,, & \qquad K[\cE_4] = -[\cC_3] \,, \\
    K[\cE_5] = - [\mathbf{E}_0] & +  [\cC_1]  + [\textcolor{red}{\cC_2}] - [\cC_4] \,, \\
    K[\cE_6] =  [\mathbf{E}_0] & - [\textcolor{red}{\cC_2}] + [\text{pt}] \,.
\end{split}
\end{align}
Its quiver diagram is drawn in Figure \ref{fig:dp3-quiver}.

The next theory we discuss is $Y^{3,q}$ family,
which has $D=5$ $\cN=1$  $SU(3)$ theory description.
The integer label $q$ is interpreted as the Chern-Simons level of the gauge theory.
Two examples of the probe quiver are illustrated in Figure \ref{fig:yp=3-quiver} in Section \ref{Q}.
$Y^{3,0}$ theory has the map between internal cycles
and fractional branes as follows.
\begin{align} \label{y30-frac-brane}
\begin{split}
    K[\cE_1] = [\mathbf{E}_1 ] \,, & \qquad K[\cE_2] = - [\mathbf{E}_1] + [\cC_3] \,, \\
    K[\cE_3] = [\mathbf{E}_2] \,, & \qquad K[\cE_4] = - [ \mathbf{E}_2 ] + [\cC_7]  \,, \\
    K[\cE_5] = - [\mathbf{E}_1] & -  [\mathbf{E}_2]  - [\cC_3] - [\cC_7] + [\textcolor{red}{\cC_9}]   \,, \\
    K[\cE_6] =  [\mathbf{E}_1] & + [\mathbf{E}_2] - [\textcolor{red}{\cC_9}] + [\text{pt}] \,.
\end{split}
\end{align}
Another interesting theory of this class is $Y^{3,1}$,
which engineers $D=5$ $SU(3)_1$ gauge theory.
\begin{align} \label{y31-frac-brane}
\begin{split}
    K[\cE_1] = [\mathbf{E}_1 ] \,, & \qquad K[\cE_2] = - [\mathbf{E}_1] + [\cC_3] \,, \\
    K[\cE_3] = [\mathbf{E}_2] \,, & \qquad K[\cE_4] = - [ \mathbf{E}_2 ] + [\cC_7]  \,, \\
    K[\cE_5] = - [\mathbf{E}_1] & -  [\mathbf{E}_2]  - [\cC_3] - [\cC_7] + [\textcolor{red}{\cC_9}]   \,, \\
    K[\cE_6] =  [\mathbf{E}_1] & + [\mathbf{E}_2] - [\textcolor{red}{\cC_9}] + [\text{pt}] \,.
\end{split}
\end{align}



\end{document}